%% file: specparams.tex
\documentclass{emulateapj}
\pdfoutput=1
\usepackage{graphicx}
\usepackage{epsfig,hyperref}
\usepackage{amssymb,amsmath}
\usepackage{array}
\usepackage{bm}              
\usepackage{natbib}
\usepackage{mathrsfs}
\usepackage{rotating}
\usepackage{afterpage}

\singlespace

\newcommand{\planck}{{\it Planck}}
\newcommand{\wmap}{{\it WMAP}}

\newcommand{\ba}{\begin{eqnarray}}
\newcommand{\ea}{\end{eqnarray}}

\newcommand{\uKrs}   {\mbox{$\mu{\rm K}\sqrt{\rm s}$}}

\newcommand  \beq    {\begin{equation}}

\newcommand  \eeq    {\end{equation}}
\newcommand  \gtsim  {\lower.5ex\hbox{$\; \buildrel > \over \sim \;$}}
\newcommand  \ltsim  {\lower.5ex\hbox{$\; \buildrel < \over \sim \;$}}

\newcommand{\LCDM}   {$\Lambda$CDM}

\newcommand{\be}{\begin{equation}}
\newcommand{\ee}{\end{equation}}
\newcommand{\mat}[1]{\ensuremath{\mathbf #1}}   
\newcommand{\neff}  {$N_{\rm eff}$}

\def\pbear{{\textsc{Polarbear}}}

\usepackage{color}
\definecolor{orange}{rgb}{1,0.3,0}

\shortauthors{}
\shorttitle{ACTPol spectrum and parameters}
\slugcomment{Draft: \today}

\begin{document}

\title{The Atacama Cosmology Telescope: Two-season ACTPol Spectra and Parameters}

\author{
Thibaut~Louis\altaffilmark{1,2},
Emily~Grace\altaffilmark{3},
Matthew~Hasselfield\altaffilmark{4,5},
Marius~Lungu\altaffilmark{6},
Lo\"ic~Maurin\altaffilmark{7},
\\
Graeme~E.~Addison\altaffilmark{8},
Peter~A.~R.~Ade\altaffilmark{9},
Simone Aiola\altaffilmark{10,11,13},
Rupert~Allison\altaffilmark{2},
Mandana~Amiri\altaffilmark{12},
Elio~Angile\altaffilmark{6},
Nicholas~Battaglia\altaffilmark{13},
James~A.~Beall\altaffilmark{14},
Francesco~de~Bernardis\altaffilmark{15},
J~Richard~Bond\altaffilmark{16},
Joe~Britton\altaffilmark{14},
Erminia~Calabrese\altaffilmark{2},
Hsiao-mei~Cho\altaffilmark{14},
Steve~K.~Choi\altaffilmark{3},
Kevin~Coughlin\altaffilmark{17},
Devin~Crichton\altaffilmark{8},
Kevin~Crowley\altaffilmark{3},
Rahul~Datta\altaffilmark{17},
Mark~J.~Devlin\altaffilmark{6},
Simon~R.~Dicker\altaffilmark{6},
Joanna~Dunkley\altaffilmark{2,3,13},
Rolando~D\"{u}nner\altaffilmark{7},
Simone~Ferraro\altaffilmark{18},
Anna~E.~Fox\altaffilmark{14},
Patricio~Gallardo\altaffilmark{15,19},
Megan~Gralla\altaffilmark{20},
Mark~Halpern\altaffilmark{12},
Shawn~Henderson\altaffilmark{15},
J.~Colin~Hill\altaffilmark{21},
Gene~C.~Hilton\altaffilmark{14},
Matt~Hilton\altaffilmark{22},
Adam~D.~Hincks\altaffilmark{12,23},
Ren\'ee~Hlozek\altaffilmark{24},
S.P. Patty~Ho\altaffilmark{3},
Zhiqi~Huang\altaffilmark{16},
Johannes~Hubmayr\altaffilmark{14},
Kevin~M.~Huffenberger\altaffilmark{25},
John~P.~Hughes\altaffilmark{26},
Leopoldo~Infante\altaffilmark{19},
Kent~Irwin\altaffilmark{27},
Simon~Muya~Kasanda\altaffilmark{22,28},
Jeff~Klein\altaffilmark{6},
Brian~Koopman\altaffilmark{15},
Arthur~Kosowsky\altaffilmark{10,11},
Dale~Li\altaffilmark{14},
Mathew~Madhavacheril\altaffilmark{29},
Tobias~A.~Marriage\altaffilmark{8},
Jeff~McMahon\altaffilmark{17},
Felipe~Menanteau\altaffilmark{30,31},
Kavilan~Moodley\altaffilmark{22},
Charles~Munson\altaffilmark{17},
Sigurd~Naess\altaffilmark{2},
Federico~Nati\altaffilmark{6,32},
Laura~Newburgh\altaffilmark{24},
John~Nibarger\altaffilmark{14},
Michael~D.~Niemack\altaffilmark{15},
Michael~R.~Nolta\altaffilmark{16},
Carolina Nu\~{n}ez\altaffilmark{13,19,33},
Lyman~A.~Page\altaffilmark{3},
Christine~Pappas\altaffilmark{3},
Bruce~Partridge\altaffilmark{34},
Felipe~Rojas\altaffilmark{19,32},
Emmanuel Schaan\altaffilmark{13},
Benjamin~L.~Schmitt\altaffilmark{6},
Neelima~Sehgal\altaffilmark{29},
Blake~D.~Sherwin\altaffilmark{18},
Jon~Sievers\altaffilmark{28,35},
Sara~Simon\altaffilmark{3},
David~N.~Spergel\altaffilmark{13,36},
Suzanne~T.~Staggs\altaffilmark{3},
Eric~R.~Switzer\altaffilmark{16,37},
Robert~Thornton\altaffilmark{6,38},
Hy~Trac\altaffilmark{33},
Jesse~Treu\altaffilmark{3},
Carole~Tucker\altaffilmark{9},
Alexander~Van~Engelen\altaffilmark{16},
Jonathan~T.~Ward\altaffilmark{6},
Edward~J.~Wollack\altaffilmark{37}
}

\altaffiltext{1}{UPMC Univ Paris 06, UMR7095, Institut d'Astrophysique de Paris, F-75014, Paris, France}
\altaffiltext{2}{Sub-Department of Astrophysics, University of Oxford, Keble Road, Oxford, UK OX1 3RH}
\altaffiltext{3}{Joseph Henry Laboratories of Physics, Jadwin Hall, Princeton University, Princeton, NJ, USA 08544}
\altaffiltext{4}{Department of Astronomy and Astrophysics, The Pennsylvania State University, University Park, PA 16802}
\altaffiltext{5}{Institute for Gravitation and the Cosmos, The Pennsylvania State University, University Park, PA 16802}
\altaffiltext{6}{Department of Physics and Astronomy, University of Pennsylvania, 209 South 33rd Street, Philadelphia, PA, USA 19104}
\altaffiltext{7}{Instituto de Astrof\'isica and Centro de Astro-Ingenier\'ia, Facultad de F\'isica, Pontificia Universidad Cat\'olica de Chile, Av. Vicu\~na Mackenna 4860, 7820436 Macul, Santiago, Chile}
\altaffiltext{8}{Dept. of Physics and Astronomy, The Johns Hopkins University, 3400 N. Charles St., Baltimore, MD, USA 21218-2686}
\altaffiltext{9}{School of Physics and Astronomy, Cardiff University, The Parade,  Cardiff, Wales, UK CF24 3AA}
\altaffiltext{10}{Department of Physics and Astronomy, University of Pittsburgh,  Pittsburgh, PA, USA 15260}
\altaffiltext{11}{Pittsburgh Particle Physics, Astrophysics, and Cosmology Center, University of Pittsburgh, Pittsburgh PA 15260}
\altaffiltext{12}{Department of Physics and Astronomy, University of British Columbia, Vancouver, BC, Canada V6T 1Z4}
\altaffiltext{13}{Department of Astrophysical Sciences, Peyton Hall,  Princeton University, Princeton, NJ USA 08544}
\altaffiltext{14}{NIST Quantum Devices Group, 325 Broadway Mailcode 817.03, Boulder, CO, USA 80305}
\altaffiltext{15}{Department of Physics, Cornell University, Ithaca, NY, USA 14853}
\altaffiltext{16}{Canadian Institute for Theoretical Astrophysics, University of Toronto, Toronto, ON, Canada M5S 3H8}
\altaffiltext{17}{Department of Physics, University of Michigan, Ann Arbor, USA 48103}
\altaffiltext{18}{Berkeley Center for Cosmological Physics, LBL and Department of Physics, University of California, Berkeley, CA, USA 94720}
\altaffiltext{19}{Departamento de Astronom{\'{i}}a y Astrof{\'{i}}sica, Pontific\'{i}a Universidad Cat\'{o}lica, Casilla 306, Santiago 22, Chile}
\altaffiltext{20}{Steward Observatory, University of Arizona.  933 North Cherry Avenue, Tucson, AZ 85721}
\altaffiltext{21}{Dept. of Astronomy, Pupin Hall, Columbia University, New York, NY 10027 USA}
\altaffiltext{22}{Astrophysics and Cosmology Research Unit, School of Mathematics, Statistics and Computer Science, University of KwaZulu-Natal, Durban 4041, South Africa}
\altaffiltext{23}{Department of Physics, University of Rome "La Sapienza", Piazzale Aldo Moro 5, I-00185 Rome, Italy.}
\altaffiltext{24}{Dunlap Institute, University of Toronto, 50 St. George St., Toronto, ON, Canada M5S3H4}
\altaffiltext{25}{Department of Physics, Florida State University, Tallahassee FL, USA 32306}
\altaffiltext{26}{Department of Physics and Astronomy, Rutgers, The State University of New Jersey, Piscataway, NJ USA 08854-8019}
\altaffiltext{27}{Department of Physics, Stanford University, Stanford, CA,  USA 94305-4085}
\altaffiltext{28}{Astrophysics and Cosmology Research Unit, School of Chemistry and Physics, University of KwaZulu-Natal, Durban 4041, South Africa}
\altaffiltext{29}{Physics and Astronomy Department, Stony Brook University, Stony Brook, NY USA 11794}
\altaffiltext{30}{National Center for Supercomputing Applications (NCSA), University of Illinois at Urbana-Champaign, 1205 W. Clark St., Urbana, IL, USA, 61801}
\altaffiltext{31}{Department of Astronomy, University of Illinois at Urbana-Champaign, W. Green Street, Urbana, IL, USA, 61801}
\altaffiltext{32}{Sociedad Radiosky Asesor\'{i}as de Ingenier\'{i}a Limitada Lincoy\'{a}n 54, Depto 805 Concepci\'{o}n, Chile}
\altaffiltext{33}{McWilliams Center for Cosmology, Carnegie Mellon University, Department of Physics, 5000 Forbes Ave., Pittsburgh PA, USA, 15213}
\altaffiltext{34}{Department of Physics and Astronomy, Haverford College, Haverford, PA, USA 19041}
\altaffiltext{35}{National Institute for Theoretical Physics (NITheP), KZN node, Durban 4001, South Africa}
\altaffiltext{36}{Center for Computational Astrophysics, 162 5th Ave NY NY 10003 USA}
\altaffiltext{37}{NASA/Goddard Space Flight Center, Greenbelt, MD, USA 20771}
\altaffiltext{38}{Department of Physics , West Chester University of Pennsylvania, West Chester, PA, USA 19383}

\begin{abstract}
We present the temperature and polarization angular power spectra measured by the Atacama Cosmology Telescope Polarimeter (ACTPol). We analyze night-time data collected during 2013--14 using two detector arrays at 149~GHz, from 548 deg$^2$ of sky on the celestial equator. We use these spectra, and the spectra measured with the MBAC camera on ACT from 2008--10, in combination with \planck\ and \wmap\ data to estimate cosmological parameters from the temperature, polarization, and temperature-polarization cross-correlations. We find the new ACTPol data to be consistent with the \LCDM\ model. The ACTPol temperature-polarization cross-spectrum now provides stronger constraints on multiple parameters than the ACTPol temperature spectrum, including the baryon density, the acoustic peak angular scale, and the derived Hubble constant.  Adding the new data to \planck\ temperature data tightens the limits on damping tail parameters, for example reducing the joint uncertainty on the number of neutrino species and the primordial helium fraction by 20\%. 
\end{abstract}

\input intro
\setcounter{footnote}{0} 

\input data

\input spectra

\input likelihood

\input params

\input conclude

\acknowledgments

This work was supported by the U.S. National Science Foundation through awards
AST-1440226, AST-0965625 and AST-0408698 for the ACT project, as well as awards PHY-1214379
and PHY-0855887. Funding was also provided by Princeton University, the
University of Pennsylvania, and a Canada Foundation for Innovation (CFI) award
to UBC. ACT operates in the Parque Astron\'omico Atacama in northern Chile
under the auspices of the Comisi\'on Nacional de Investigaci\'on Cient\'ifica y
Tecnol\'ogica de Chile (CONICYT). Computations were performed on the GPC
supercomputer at the SciNet HPC Consortium and on the hippo cluster at the University of KwaZulu-Natal. SciNet is funded by the CFI under
the auspices of Compute Canada, the Government of Ontario, the Ontario Research
Fund -- Research Excellence; and the University of Toronto. The development of
multichroic detectors and lenses was supported by NASA grants NNX13AE56G and
NNX14AB58G. CM acknowledges support from NASA grant NNX12AM32H. JD and SN are supported by ERC grant 259505. EC is supported by a STFC Rutherford Fellowship. TL  is supported by ERC grant 267117, by ERC grant 259505, and by the Labex ILP (reference ANR-10-LABX-63) part of the Idex SUPER, and received financial state aid from ANR-11-IDEX-0004-02. HT is supported by grants NASA
ATP NNX14AB57G, DOE DE-SC0011114, and NSF AST-1312991. AK has been supported by grant NSF AST-1312380.
BS, BK, CM, and EG are funded by NASA Space Technology Research Fellowships.
R.D received funding from the CONICYT grants FONDECYT-1141113, PIA Anillo ACT-1417 and BASAL PFB-06 CATA. LM is funded by ALMA-CONICYT grant 31140004.
We thank our many
colleagues from ABS, ALMA, APEX, and \pbear{} who have helped us at critical
junctures. Colleagues at AstroNorte and RadioSky provide logistical
support and keep operations in Chile running smoothly.
We also thank the Mishrahi Fund and the Wilkinson Fund for their generous
support of the project.

\bibliographystyle{act}
\bibliography{all}

\appendix 

\section{ACTPol Two-season Power Spectra}
\input spectra_tables
\section{Coaddition of spectra and Covariance matrix}

The power spectra $C^{XY}_{\ell}$ ($X,Y=\{T,E,B \}$) presented in this paper are the result of a weighted average of nine data spectra, \ba \{ d^{PA1}_{56}\times d^{PA1}_{56}, d^{PA1}_{56}\times d^{PA2}_{56},d^{PA2}_{56}\times d^{PA2}_{56}, d^{PA1}_{5}\times d^{PA1}_{5}, d^{PA1}_{6}\times d^{PA1}_{6}, d^{PA1}_{5}\times d^{PA1}_{56},d^{PA1}_{5}\times d^{PA2}_{56}  ,  d^{PA1}_{6}\times d^{PA1}_{56},d^{PA1}_{6}\times d^{PA2}_{56}.  \} \nonumber \ea Assuming a Gaussian distribution for these power spectra, their maximum likelihood combination can be easily obtained as
\ba
C^{XY, (ML)}_{\ell}= (P^{T} { \bm \Sigma}^{-1} P)^{-1} (P^{T} { \bm \Sigma}^{-1} ) {\bm C^{XY}_{\ell}},
\ea
where ${ \bm {C}^{XY}_{\ell}}$ is a vector encompassing all data spectra,  $ { \bm \Sigma}$ is the covariance matrix of all the different spectra pairs and P is a projection matrix. $P^{T}$ projects all the individual spectra into a single power spectrum. The covariance matrix of the maximum likelihood spectra is then given by
\ba
{\bm \Sigma^{(ML)}}=(P^{T} { \bm \Sigma}^{-1} P)^{-1}.
\ea

\begin{figure}[ht!]
\centering
	\includegraphics[width=\textwidth]{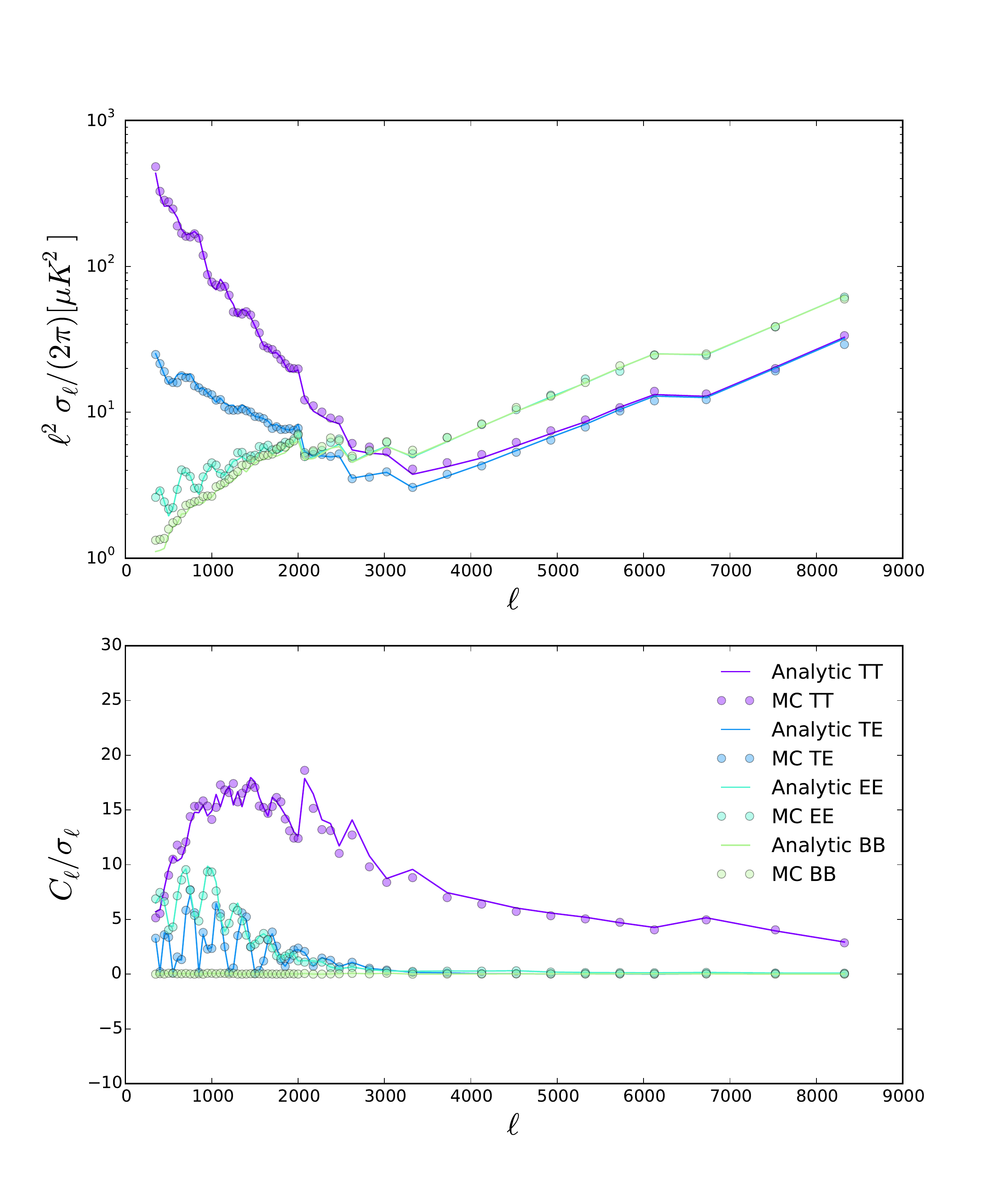}
\caption{Comparison between analytic and Monte Carlo estimates of the errors for D56 PA1. The top panel shows the errors and the bottom panel shows the signal to noise on the TT, EE, TE and BB D56 PA1 spectra.}
\label{fig:errors}
\end{figure}

We estimate the covariance matrix of all the spectra using 840 Monte Carlo simulations, with
\ba
 \Sigma^{(\alpha A \times \beta B);( \gamma C \times \tau D)}_{\textrm{WXYZ, bb}}  =  \langle (\hat{C}_{b,WX} ^{(\alpha A \times \beta B)}-  \langle \hat{C}_{b,WX} ^{(\alpha A \times \beta B)} \rangle) (\hat{C}_{b,YZ} ^{( \gamma C \times \tau D)}-  \langle \hat{C}_{b,YZ} ^{(\gamma C \times \tau D)} \rangle) \rangle. \nonumber
\ea
Here A,B,C,D stands for detector arrays (PA1 and PA2), $\alpha$,$\beta$,$\gamma$,$\delta$ stands for season of observation and W,X,Y,Z stands for T, E, B. 
We use this estimate in the likelihood, and also verify that the dispersion of the simulation is consistent with an analytical estimate:
\ba
\Sigma^{(\alpha A \times \beta B);( \gamma C \times \tau D)}_{\textrm{WXYZ, bb}} &=& \frac{1}{\nu_b}\left( S_{b,WY} S_{b,XZ} +S_{b,WZ} S_{b,XY} \right)   \nonumber \\
&+&\frac{1}{N_{s} \nu_b}  \left[S_{b,WY}\delta_{\beta \tau} N_{b,XZ} ^{ \beta B \times \tau D}+S_{b,XZ}\delta_{\alpha \gamma} N_{b,WY} ^{ \alpha A \times  \gamma C} +  S_{b,WZ}\delta_{\beta \gamma} N_{b,XY} ^{  \beta B \times  \gamma C}  +S_{b,XY}\delta_{\alpha \tau} N_{b,WZ} ^{ \alpha A \times \tau D}  \right] \nonumber \\
&+& \frac{1}{\nu_b}\frac{N_{s}^{2}-N_{s}(\delta_{\alpha \beta}+\delta_{\gamma \tau})+N_{s}\delta_{\alpha \beta}\delta_{\gamma \tau}}{N^{4}_{s}-N^{3}_{s}(\delta_{\alpha \beta}+\delta_{\gamma\tau})+ N^{2}_{s}\delta_{\alpha \beta}\delta_{\gamma \tau}}\left[ \delta_{\alpha \gamma}  \delta_{\beta \tau}  N_{b,WY} ^{\alpha A \times \gamma C} N_{b,XZ} ^{ \beta B \times \tau D}+\delta_{\beta \gamma}\delta_{\alpha \tau} N_{b,WZ} ^{\alpha A \times \tau D}  N_{b,XY} ^{ \beta B \times \gamma C}  \right]. \nonumber
\ea
Here, S is the signal power spectrum, N represents the noise power spectrum, $\nu_b$ is the number of modes in the bin b, and $N_{s}$ is the number of splits used for computing each cross spectra.
The analytic and Monte Carlo estimates are shown in Figure  \ref{fig:errors}.
The final covariance matrix is then given by the sum of the signal and noise covariance matrix, the calibration covariance matrix, the beam covariance matrix and a covariance matrix accounting for possible residual leakage estimated using planet map observations
\ba
{ \bm \Sigma^{\textrm{all}}}= { \bm \Sigma}+ { \bm \Sigma^{\textrm{cal}}} + { \bm \Sigma^{\textrm{beam}}} +  { \bm \Sigma^{\textrm{leakage}}}. 
\ea

\end{document}

%% file: intro.tex
\section{Introduction}
\label{sec:intro}

The now standard $\Lambda$CDM model of cosmology has been increasingly refined with measurements of the cosmic microwave background (CMB), most recently by the \planck\ satellite \citep{planck_params/2013, planck_cosmo:2015}. This model provides an excellent fit to current cosmological data but leaves unanswered questions about the contents, structure and dynamics of the Universe, and their origins. Some tensions exist at the 2-3$\sigma$ significance level between the Hubble constant and the amplitude of fluctuations derived from different cosmological probes \citep[e.g.,][]{riess/etal:2016,hildebrandt/etal:2016}. One of the paths forward is an improved measurement of the polarization anisotropy and its power spectra.

Significant new CMB polarization data have been published in the last three years. The \planck\ team reports TE and EE polarization spectra for $\ell \geq 50$ from the HFI instrument \citep{planck_cosmo:2015}, and estimates the large-scale E-mode signal from the LFI and HFI instruments \citep{planck_like:2015,planck_reion:2016}. The E-mode power spectrum has also been measured by \wmap\ on large scales \citep{hinshaw/etal:2013}, and on smaller scales with first-season ACTPol data \citep{naess/etal:2014}, by BICEP2/Keck \citep{2016PhRvL.116c1302B}, \citet{pbear-eebb/2014}, and SPTpol \citep{2015ApJ...805...36C}. These all show the E-mode signal to be consistent with the $\Lambda$CDM prediction. 

The B-mode gravitational lensing signal has now been measured at 2$\sigma$ by the \citet{pbear-eebb/2014}, at 4$\sigma$ by SPTpol \citep{keisler/etal:2015}, and at 7$\sigma$ by BICEP2/Keck \citep{2016PhRvL.116c1302B}. It has been detected in cross-correlation with the reconstructed lensing signal by SPTpol \citep{hanson/etal/2013},  The Polarbear Collaboration: \citet{pbear-eeeb/2013}, ACTPol \citep{2015ApJ...808....7V}, and \planck\ \citep{planck_lens:2015}. 

This paper describes the temperature and polarization power spectra and derived cosmological parameters obtained from two seasons of observations by the Atacama Cosmology Telescope Polarimeter (ACTPol). In this analysis we use only data collected at night in a 548 deg$^2$ region known as `D56.' In \S2 we describe the data and basic processing, and in \S3 show the power spectra and null tests. In \S4 we describe our likelihood method, in \S5 show cosmological results, and conclude in \S6.

%% file: data.tex
\section{Data and processing}
\label{sec:data}

In this paper we use a combination of data collected during three months of observations in 2013 using a single detector array known as PA1, as reported in Naess et al. (2014), combined with data from a four month period in 2014 using the PA1 and PA2 detector arrays.
Each detector array is coupled to 522 feedhorns, and has 1044 TES bolometers operating at 149 GHz, of which a median 400 (for PA1) and 600 (for PA2) detectors are used for this analysis. Further description of the instrument is given in \citet{naess/etal:2014} and \citet{thornton/etal:prep}. 

We refer to the first-season 2013 data as S1, and the second-season 2014 data as S2.  Observations of Uranus permit the direct calibration of timestream data to estimate detector sensitivities.  These measurements produce array noise equivalent temperatures (NETs) of 15.3~\uKrs\ and 23.0~\uKrs\ for PA1 in S1 and S2 respectively, and 12.9~\uKrs\ for PA2.  The sensitivities of the detector arrays depend on the loading from the sky. These values correspond to a precipitable water vapor column density, along the line of sight, of 1.2 mm, which was the median value for S2 observations.  The decreased sensitivity of PA1 in S2 is due to higher average cryogenic temperatures of the detectors. Because of data cuts, the white noise levels seen in the CMB maps are 12$\%$ higher, in temperature, than the simple prediction based on these array sensitivities and the observing time.

The passbands for both PA1 and PA2 detectors were measured in the field using a Fourier Transform Spectrometer coupled to the cold optics at the receiver windows. The effective frequency for the CMB is $\nu_{\rm CMB} = 148.9 \pm 2.4$~GHz for PA1 and $\nu_{\rm CMB} = 149.1 \pm 2.4$~GHz for PA2 \citep{thornton/etal:prep}.

\begin{figure*}
	\vspace*{-10mm}
	\hspace*{-7mm}\begin{tabular}{m{5mm}m{18cm}}
		H & \includegraphics[width=18cm,clip,trim={0mm 11mm 0mm  0mm}]{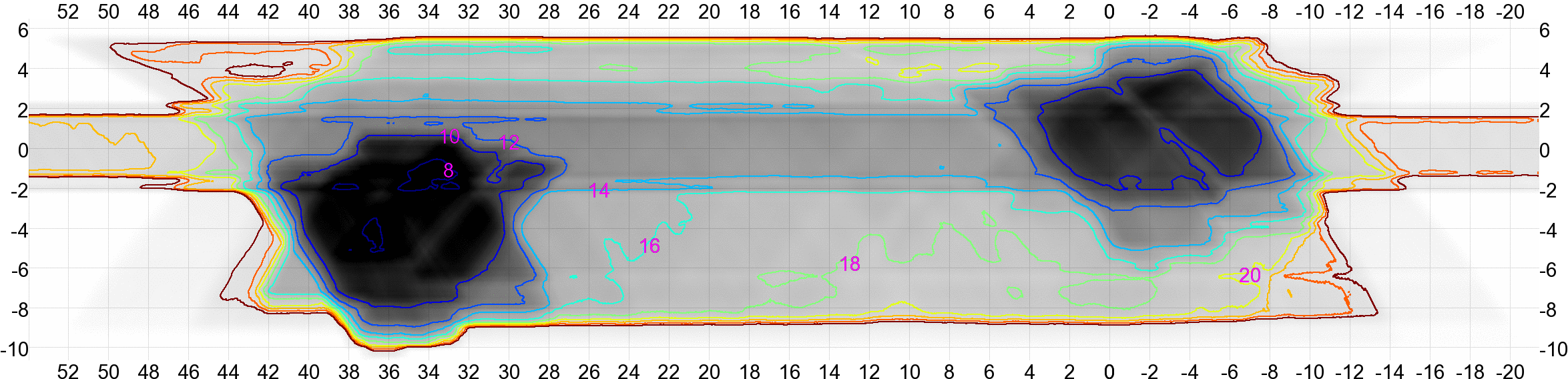}\\
		T & \includegraphics[width=18cm,clip,trim={0mm 11mm 0mm 11.5mm}]{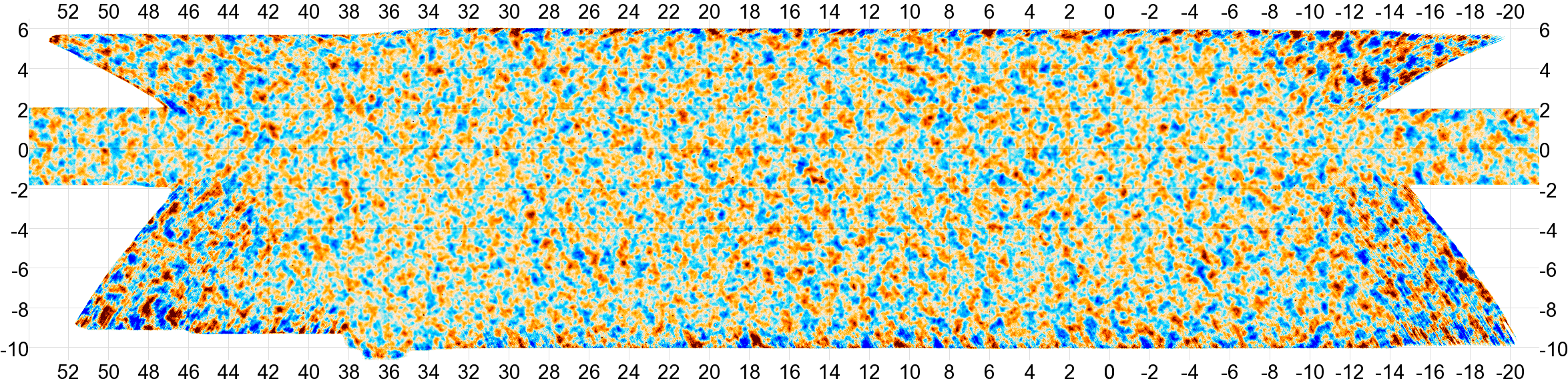}\\
		Q & \includegraphics[width=18cm,clip,trim={0mm 11mm 0mm 11.5mm}]{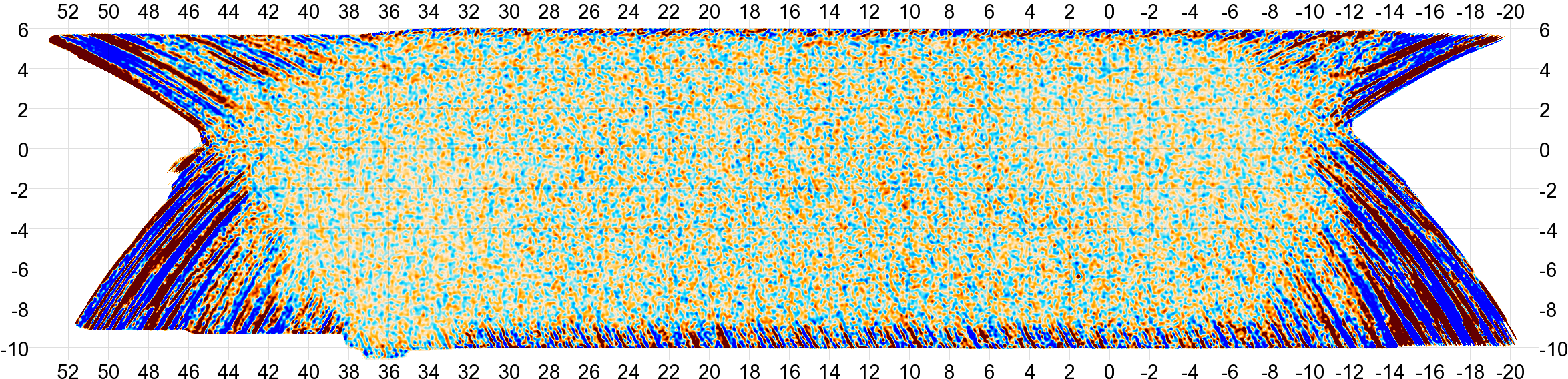}\\
		U & \includegraphics[width=18cm,clip,trim={0mm 11mm 0mm 11.5mm}]{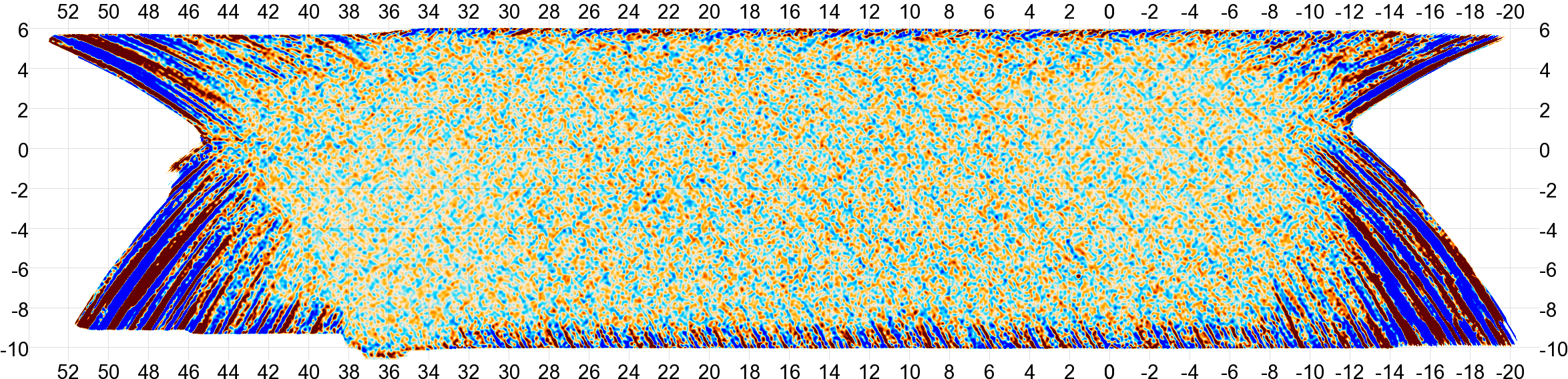}\\
		E & \includegraphics[width=18cm,clip,trim={0mm 11mm 0mm 11.5mm}]{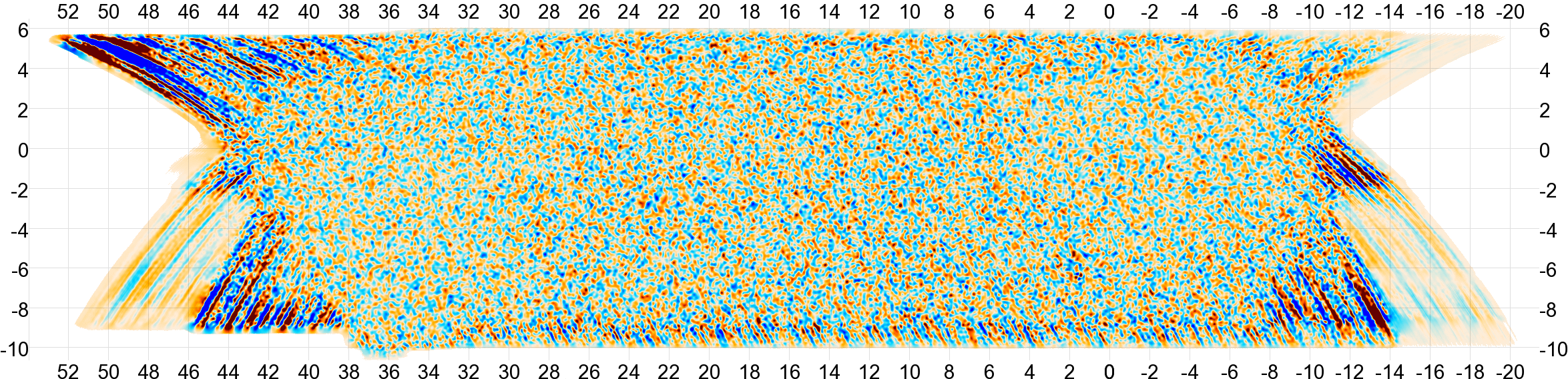}\\
		B & \includegraphics[width=18cm,clip,trim={0mm  0mm 0mm 11.5mm}]{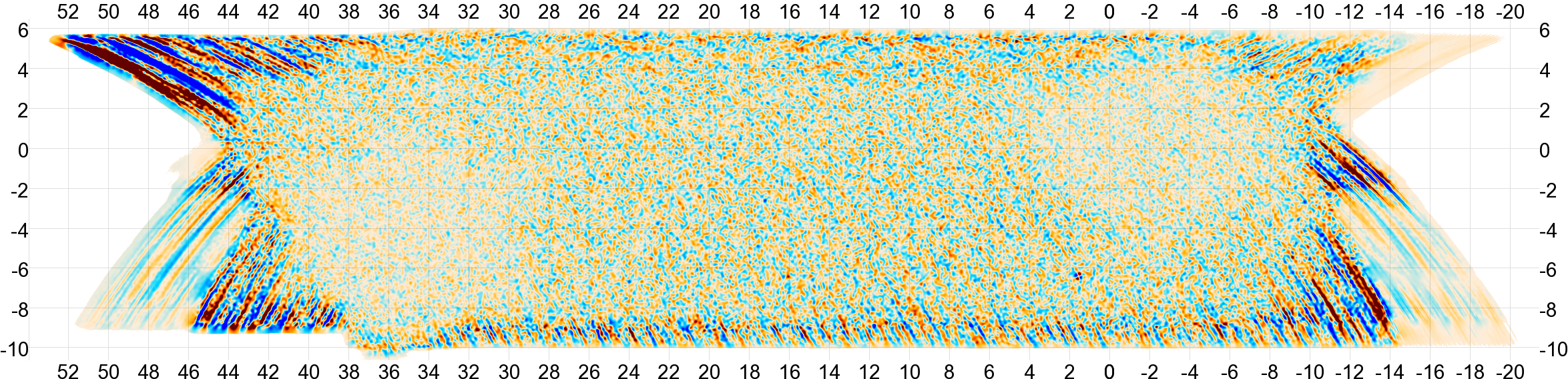}\\
	\end{tabular}
	\caption{Top (H): Exposure map in equatorial coordinates (the horizontal and vertical axes are RA and Dec respectively), including both the three-season MBAC data and the ACTPol data used in this analysis. The D5 and D6 regions are the deep fields on the right and left sides of the map, and D56 is the wider rectangle which overlaps both deep fields. The contour labels indicate the T noise level in
	$\mu\textrm{K}\cdot\textrm{arcmin}$, starting from $8\mu\textrm{K}\cdot\textrm{arcmin}$
	in the deepest region. The Q and U noise levels are each $\sqrt{2}$ higher. Lower
	panels: Filtered maps in T and in Q, U, E and B-polarization. All maps are
	filtered with a highpass-filter at $\ell=200$ and a horizontal highpass-filter at
	$\ell=40$. The polarization maps are additionally lowpass-filtered at $\ell=1900$.
	The color scale is $\pm250\mu$K in T and $\pm25\mu$K in P.}
	\label{fig:maps}
\end{figure*}

\subsection{Observations}
In 2013 ACTPol observed four deep regions covering 260 deg$^2$ at right ascensions 150$^\circ$, 175$^\circ$, 355$^\circ$, and 35$^\circ$, known as D1, D2, D5 and D6. In the second and third seasons, ACTPol observed two wider regions, known as D56 and BOSS-N. The D56 region used for analysis covers 548 deg$^2$ with coordinates $ -7.2^\circ<{\rm dec}<4^\circ$ and $352^\circ<{\rm RA}<41^\circ$, and BOSS-N covers 2000 deg$^2$ with coordinates $-4^\circ<{\rm dec}<20^\circ$ and $142^\circ<{\rm RA}<228^\circ$. The D5 and D6 sub-regions lie within D56. The D56 and BOSS-N regions are visible to the telescope at different times of day, and each was observed both rising and setting on each day.  The observations alternated, from day to day, between two different elevations, to provide a total of four different parallactic angles in the complete data set. Data were taken from Sept. 11, 2013 to Dec. 14, 2013 (S1), and Aug. 20,  2014 to Dec. 31, 2014 (S2).

\begin{figure}
  \centering
	\includegraphics[width=3.5in]{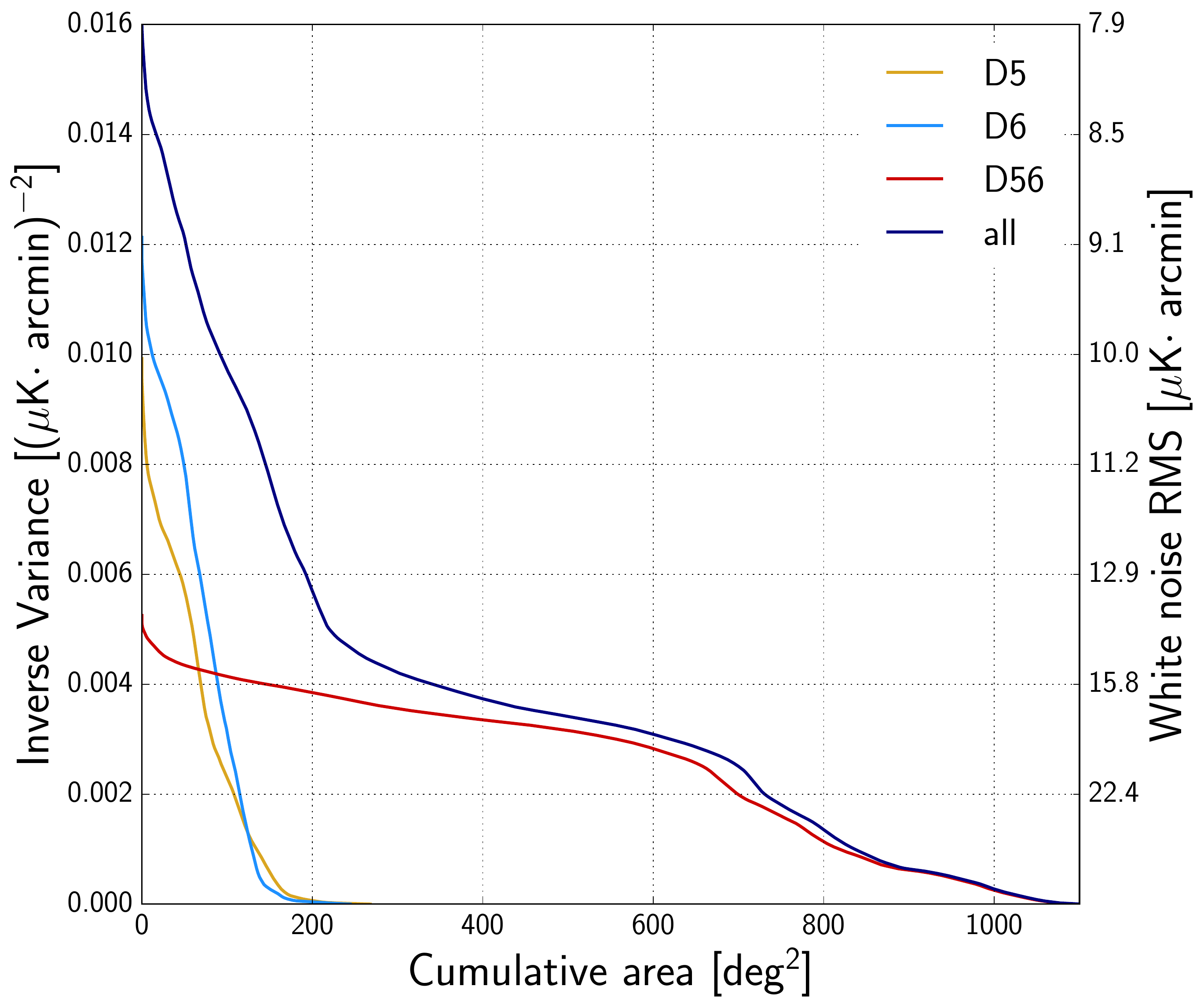}
\caption{The temperature white noise levels (right axis), and inverse variance (left axis), in the ACTPol maps as a function of cumulative area. Levels are shown for the larger D56 region, the smaller D5 and D6 sub-regions, and the combined map.\\}
\label{fig:noise_map}
\end{figure}

In this paper we analyze just the night-time data in the D56 region, including the D5 and D6 sub-regions measured in S1.
These data correspond to 45\% of all two-season CMB data that pass data quality screening procedures (55\% of S1 and 40\% of S2), and 12\% of all screened three-season data.
The combined maps and weight map of the two-season data are shown in Figure \ref{fig:maps} and a summary of the data given in Table \ref{tab:data}.  As in \citet{naess/etal:2014} we analyze only the lowest noise regions of the maps. Combining the data from PA1 and PA2 for D56, and additionally including S1 data for D5 and D6, this results in a white noise map sensitivity of 18, 12, and 11 $\mu$K$\cdot$ arcmin for D56, D5 and D6 respectively, illustrated in Figure \ref{fig:noise_map}. To get Stokes Q or U sensitivities, multiply by $\sqrt{2}$.  

\begin{table}[t!]
\caption{\small Summary of two-season ACTPol data used in this analysis: night-time data in D56 region.}
\centering
\begin{tabular}{lccc}
\hline
\hline
     & D56   &   D5 & D6 \\
\hline
RA min, max(deg)   & $-8.0, 41.0$  & $-7.5, 2.7$ & $30.0, 40.0$\\
Dec min, max (deg)  &  $-7.2, 4.0$ &  $-3.0,3.8$  & $-7.2, -1.0$  \\
Analyzed area (deg$^2$) & 548 &  70 & 63 \\
Noise level ($\mu$K.arcmin) & 17.7 & 12.0 & 10.5\\
Hours (S1) & & 222 & 268 \\
Hours (S2) &709\footnotemark[1] &  &  \\ 
Effective $N_{\rm det}$\footnotemark[2]  & 457 & 404 & 430\\
\hline
\footnotetext[1]{Observing hours summed over the two arrays, with 337 hrs in PA1 and 372 in PA2.}
\footnotetext[2]{The total amount of data is the effective number of detectors multiplied by observing hours.}
\end{tabular}
\label{tab:data}
\end{table}

\begin{figure*}
\centering
	\includegraphics[width=13cm]{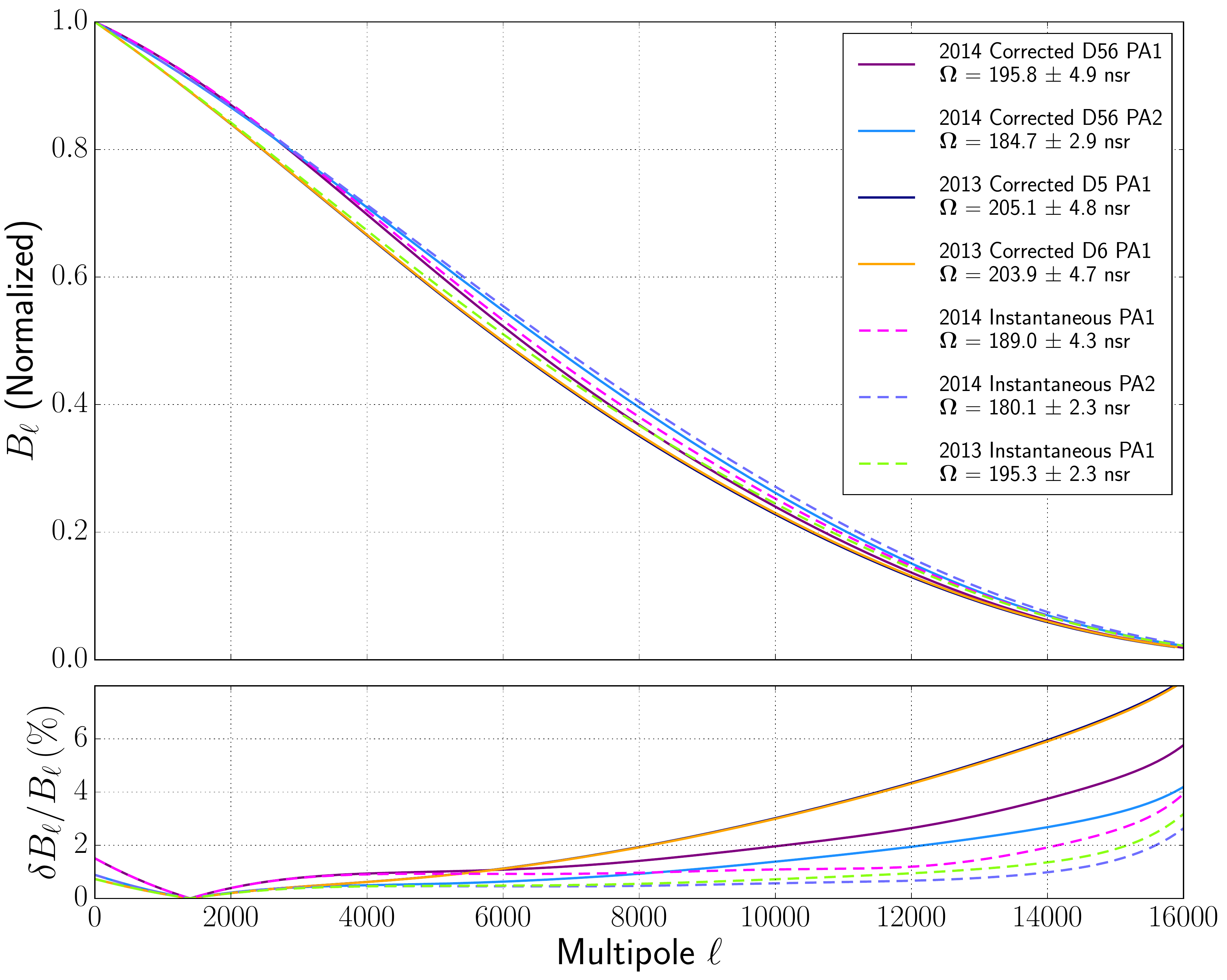}
\caption{The beam window functions (top) and uncertainties (bottom) measured by ACTPol during the first (S1, 2013) and second (S2, 2014) observing seasons for the arrays PA1 and PA2. Both the instantaneous beams (dashed lines) and the pointing variance corrected beams (solid lines) for the three different regions included in the analysis are shown. The total solid angle and its uncertainty are given for each beam in units of nanosteradians (nsr).}
\label{fig:beam}
\end{figure*}

\subsection{Data pre-processing}

Much of the data selection and analysis follows \citet{naess/etal:2014} and \citet{dunner/etal/2013}; here we report changes or improvements made for this analysis. 

The data selection method has been refined and sped up. The new algorithm works mainly in frequency space, assessing the data properties and systematics in different frequency bands. In the sub-Hz range, the data are dominated by atmosphere temperature brightness fluctuations, and to a lesser degree bath thermal drifts. The latter are measured and deprojected using the signal from detectors which are not optically coupled, known as dark detectors. The correlations generated by the atmospheric drifts are used to select the working detectors and measure their relative gains.
The detector noise is measured at higher frequencies, between 8 and 15 Hz, after deprojecting the ten largest modes across the array corresponding to up to 10\% of the total variance. Detectors with an extreme noise level, or abnormal skewness or kurtosis, which is a signature of residual contamination, are rejected from the analysis.

\subsection{Pointing and beam}

The pointing reconstruction model has been improved. The preliminary pointing model is still constrained using observations of planets at night. In this new analysis we also apply a correction to account for temporal variation in the pointing of the telescope. We first make a preliminary map with the nominal pointing estimate. We then locate bright point sources and find their true positions by matching them to known catalogs.  Assuming that the beam is constant, we then take each $\approx$10 minute section of time-ordered data, which we term a `TOD,' and perform a joint fit in the time domain to the four brightest sources stronger than 1~mK, fitting for a single overall pointing offset per TOD, $[p_x,p_y]$, in focal plane coordinates, and a flux for each source.

We use this primary pointing correction when a TOD has a good source fit, quantified by requiring the uncertainty on the pointing offset, $\sigma(p_{x,y})$ to be less than $12\arcsec$, which is satisfied for 65\% of the data. For the remaining TODs we follow a simple prescription. If available, we use the average of nearest neighbor TODs within 15 minutes, of the same scan type with the same azimuth and elevation (24\% of the data). If not available, we use secondary neighbors within 30 minutes (4.5\%). If fits from neighboring TODs are not available, we use an average of offsets from the same scan type within 0.5, 1, or 1.5 hours in UTC (5\%). If none of the above is possible, we use an average of all TODs with good fits within 0.5 hours in UTC (2\%), which amounts to correcting for a global offset. Maps are remade with this refined pointing solution.

As in the first season, we use multiple observations of Uranus to determine the beam profile, which is modeled in one radial dimension. The beam window functions and solid angles are described in \citet{thornton/etal:prep}, and are normalized at $\ell=1400$. 
The beam uncertainty is further increased due to the position-dependent pointing uncertainty. The impact of this uncertainty on the full season maps is handled by projecting the estimated pointing variance for each TOD, weighted by white noise level, into a map and finding its distribution. This is combined with an estimate made by convolving the instantaneous beam with a Gaussian, and fitting for its width using multiple bright point sources.
The resuting pointing variance corrected beams, along with the instantaneous beams, are shown in Figure \ref{fig:beam}. The corrected beams are included in the covariance matrix for the power spectrum, following the same treatment as in \citet{naess/etal:2014}.

\begin{figure*}
\centering
	\includegraphics[width=0.74\textwidth]{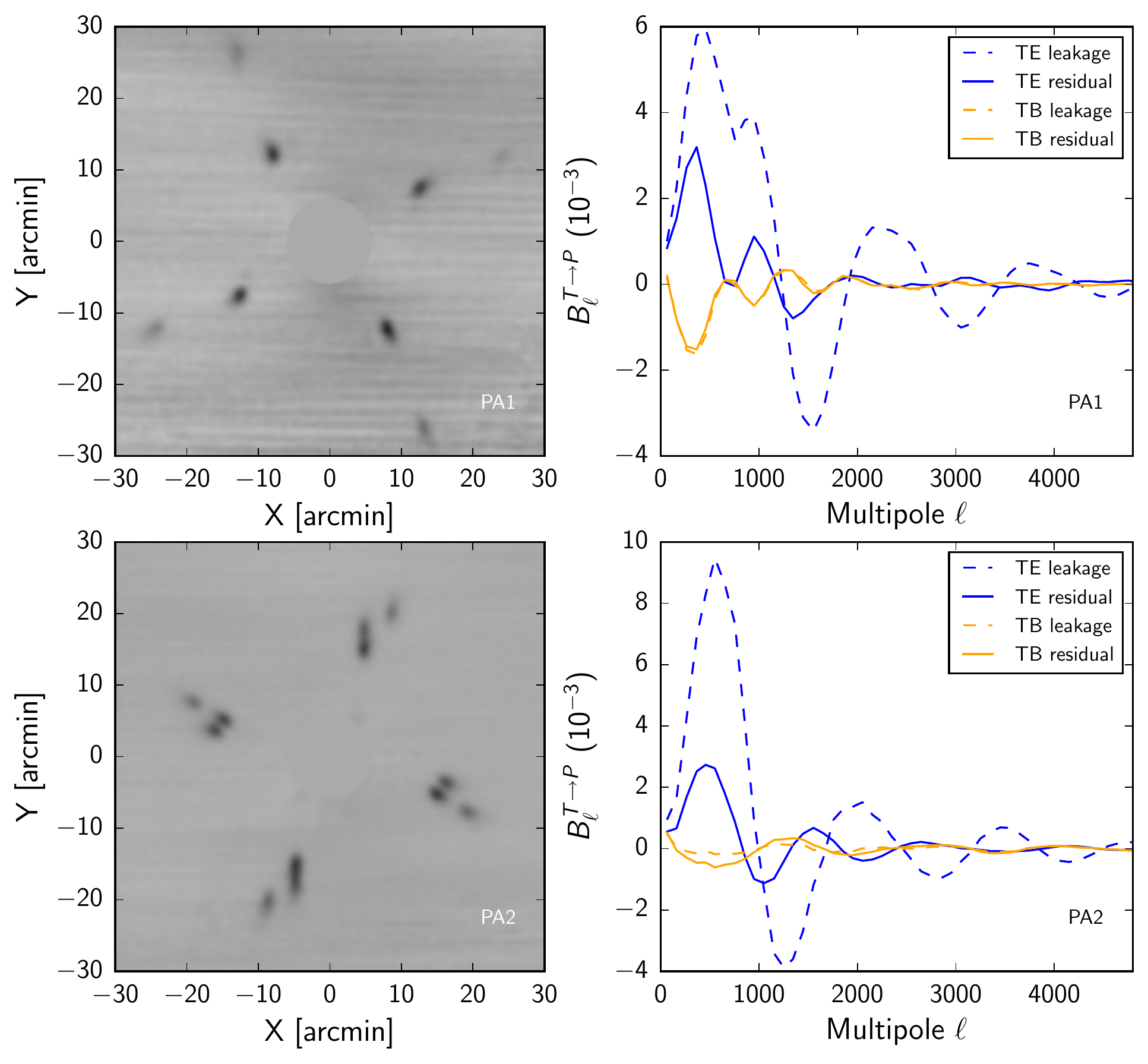}
\caption{Polarized sidelobes in PA1 (top row) and PA2 (bottom row).  Left panels show maps of the beam sidelobes, from 20 observations of Saturn.  Spatial coordinates are relative the main beam, which is masked here.  Grayscale provides the sidelobe amplitude in the range -0.002 (black) to +0.001 (white) relative to the main beam peak, with negative signal indicating polarization perpendicular to the ray from the origin.  The complementary polarization component (corresponding to TB leakage) is smaller and not shown in the maps but is included in the evaluation of the transfer functions.  Right panels show the TE and TB transfer functions, normalized in units of the main beam, as in Figure \ref{fig:beam}, before and after the sidelobe deprojection procedure.}
\label{fig:sidelobe}
\end{figure*}

During this analysis we established the existence of weak, polarized sidelobes in the PA1 and PA2 optical systems.  The sidelobes are shown in Figure \ref{fig:sidelobe} and consist of several slightly elongated images of the main beam, suppressed to a level below -30 dB and distributed with a rough 4-fold symmetry at a distance of approximately 15\arcmin (with an additional set visible at 30\arcmin in PA1).  The sidelobes are strongly polarized in the direction perpendicular to the vector between the main beam and the sidelobe position, which results in a small leakage of intensity into E-mode polarization.  

Studies of Saturn observations across the three ACTPol observing seasons show that the sidelobes are stable in time, and that their amplitudes are stable across the focal plane with the exception that there are no sidelobes associated with any points outside each array's field of view.  The fact that sidelobes from Saturn are only seen if Saturn lies within a focal plane's field of view confirms that the effect originates inside the receiver and not in the primary or secondary optics.

To remove this effect from the maps, the sidelobe signal is projected out of the time domain data prior to the mapmaking stage.  To facilitate this deprojection, the sidelobes are modeled as a combination of $\approx 20$ instances of the main beam, with the T, Q, and U amplitudes (relative to the main beam in focal plane coordinates) of each instance fitted to Saturn observations.  As a test of this removal process and to estimate residuals, we run Saturn observations through the map-making pipeline and demonstrate significant improvement in the TE and TB transfer functions (see Figure \ref{fig:sidelobe}).  The remaining TE and TB contamination is treated as a systematic error in the cosmological spectrum analysis.

The origin of these polarized sidelobes is under investigation; the optical characteristics suggest the effect is related to the filter element near the Lyot stop. We do not observe visible sidelobes in the PA3 data, which has a different configuration of filters.

As in  \citet{naess/etal:2014}, the polarization angles for the PA1 and PA2 detectors are calculated by detailed optical modeling of the mirrors, lenses and filters \citep{2016arXiv160701825K}.  A rotating wire grid was used to confirm that, apart from a global offset angle, the relative orientations of the detectors differ from the optical model with an RMS of less than two degrees.  Because the optical modeling ties together the positions of the detectors and their polarization angles, the measurement of the relative positions of the detectors on the sky also fixes the polarization angles of all detectors on the sky.  There are thus no free parameters in mapping the optical model to the sky; work is underway to quantify any remaining systematic error in the optical modeling procedure.  We later test for any additional global angle offset using the EB power spectrum.

\subsection{Mapmaking}

We continue to estimate maps using the maximum likelihood method, solving the system
\be
(A^T N^{-1} A) \bm{m}  = A^T N^{-1} \bm{d}
\ee
using the preconditioned conjugate gradient algorithm described in \citet{naess/etal:2014}. Here, $\bm{d}$ is the set of time-ordered data, A is the generalized pointing matrix that projects from map domain to time domain, and N is the noise covariance of the data. We make separate maps for the D56 wide region for both PA1 and PA2, and maps of the deeper D5 and D6 sub-regions for PA1. In each case we make four map-splits, allocating every fourth night of data to each split. The map depths are shown in Figure \ref{fig:noise_map}. As in \citet{naess/etal:2014} we use cylindrical equal area (CEA) pixels of side 0.5\arcmin, in equatorial coordinates.

We now account for the beam sidelobes in the mapmaking as described in the previous section. We also make a set of cuts in the mapmaking step. We use the same treatment to remove scan-synchronous pick-up as in \citet{naess/etal:2014}, applying an azimuth filter to the time-ordered data. We also detect several spikes in the TOD power spectra, and mask those as a precaution. 
We found a few detectors occasionally deviate from the expected white-noise behavior at high frequency. To avoid giving these unrealistically high statistical weight in the mapmaking, we apply a cut requiring the noise power at 100 Hz to be no more (less) than 3 (0.5) times the power at 10 Hz.

To identify possible systematic contaminants we make maps centered on the Moon and on the Sun, as well as in coordinates fixed relative to the ground. To remove Moon contamination we make a new Moon-centered mask defined using a Sun-centered map to better measure the beam sidelobes. This new mask reduces the number of TODs by 7\%, and includes masking sidelobes at $30^\circ$ away from the boresight that were not masked in the original S1 maps, in addition to the two sidelobes at $20^\circ$ and $120^\circ$ identified in \citet{naess/etal:2014}.

To remove ground or other scan-synchronous pickup contamination we bin
each detector's data by azimuth for each of the different scanning
patterns. Several classes of near-constant excess signal are observed
for groups of detectors at particular regions of azimuth and
elevation. Some of these we attribute to ground pick-up, but most of
them appear to be internal to the telescope. We mask these regions,
corresponding to removing 6\% of the data. These cuts will be described
more fully in a follow-up paper. Their effect on the spectra is tested
in one of our null tests.

Finally, we re-estimate the transfer function for these new maps, finding it to
decrease from 0.995 at $\ell = 500$ to 0.95 at $\ell = 200$. 

\begin{figure*}
	\vspace*{-0mm}
	\tabcolsep=0.5mm
	\hspace*{-0mm}
	\begin{tabular}{m{5mm}m{18cm}}
		\begin{turn}{90}ACTPol\end{turn} &
			\includegraphics[width=17cm,clip,trim={0.0mm 6.7mm 0.0mm 0.0mm}]{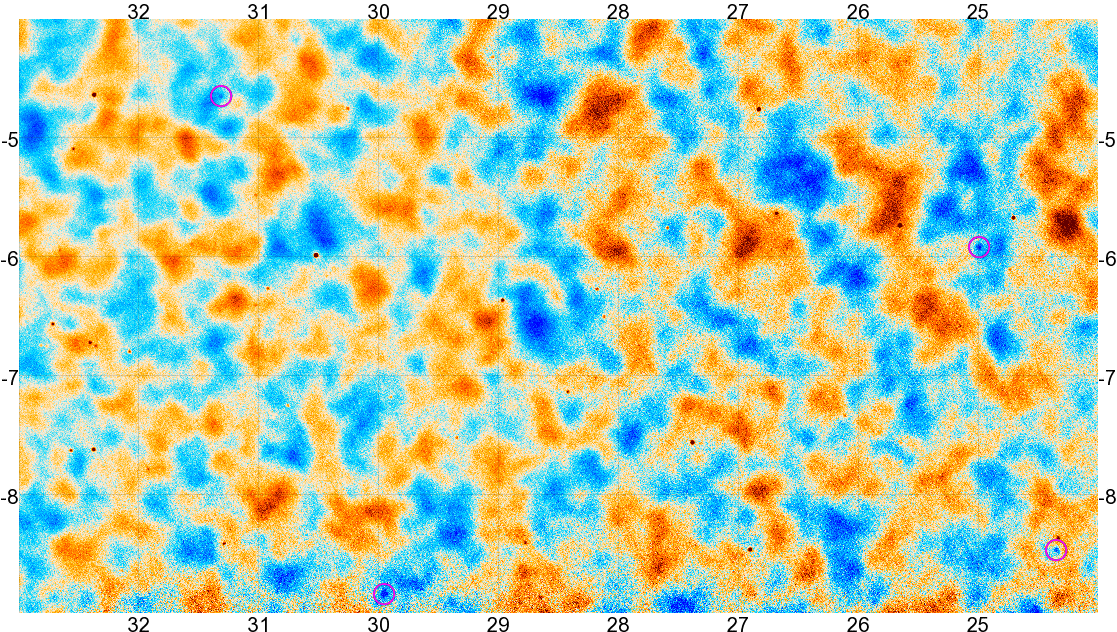} \\
		\begin{turn}{90}Planck\end{turn} &
			\includegraphics[width=17cm,clip,trim={0.0mm 0.0mm 0.0mm 6.7mm}]{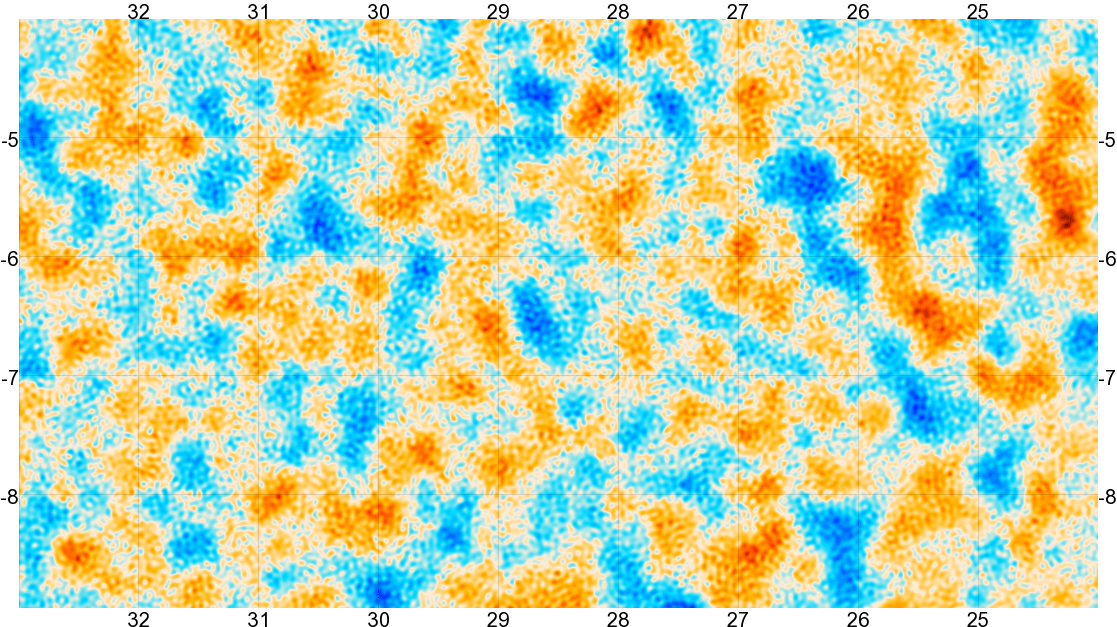} \\
	\end{tabular}
	\caption{A 45 deg$^2$ subset of the map in full resolution in T
	showing ACTPol 149 GHz (top) and
	Planck 143 GHz (bottom), in equatorial coordinates, both filtered as in Figure 1. The color scale is $\pm250\mu$K. This region covers the transition
	from deep (top left, sensitivity $10\mu\textrm{K}\cdot\textrm{arcmin}$) to shallow
	(right, $16\mu\textrm{K}\cdot\textrm{arcmin}$) exposure, and represents
	about 8\% of the usable area in D56. The two maps
	are in good agreement. Several point sources (red dots) and SZ clusters (circled) are visible in the ACTPol map. The identified clusters are ACT-CL J0137.4-0827,
ACT-CL J0140.0-0554, ACT-CL J0159.8-0849 (all
previously found in other cluster surveys), and ACT-CL J0205.3-0439 (reported in \citet{naess/etal:2014}). Their details will be given in a forthcoming paper.}
	\label{fig:zoom}
\end{figure*}

A 45 deg$^2$ cut-out of the ACTPol temperature map is shown in Figure \ref{fig:zoom}, compared to the corresponding part of the \planck\ 143 GHz map. This region  covers the transition from the deep to the shallower part of the ACTPol data. The two maps are in good agreement; a quantitative comparison of the data is presented in Section \ref{sec:spectrum}.

%% file: spectra.tex
\section{Angular power spectra}
\label{sec:spectrum}

\subsection{Methods}
\label{subsec:methods}

We follow the methods described in  \citet{louis/etal/2013} and \citet{naess/etal:2014} to compute the binned angular power spectra using the flat-sky approximation. This is a standard pseudo-$C_{\ell}$ approach that accounts for the masking and window function with a mode-coupling matrix. In this analysis we do not use the pure B-mode estimator as in \citet{smith:2006} as our focus is on E-modes. We compute cross-spectra from four map-splits, and following \citet{das/etal/2014} and \citet{naess/etal:2014} we mask Fourier modes with $|k_x|<90$ and $|k_y|<50$ to remove scan-synchronous contamination. We identify bright point sources in the intensity map, and mask 273 sources with flux brighter than 15 mJy using a circle of radius 5\arcmin. We do not mask any polarized point sources or SZ clusters, but we identify one bright polarized point source at a previously-known source location RA = 1.558$^\circ$, Dec = -6.396$^\circ$. We present power spectra in the range $500< \ell < 9000$ in temperature and $350< \ell < 9000$ in polarization, chosen to minimize atmospheric contamination, large-scale systematic contamination, and to avoid angular scales where the transfer function deviates from unity by more than a percent.

We compute the power spectra for D56 (PA1$\times$PA1, PA2$\times$PA2, and PA1$\times$PA2),  for D5 and D6 PA1$\times$PA1, and for the cross-correlation between the deep regions (D5, D6) and D56. As in \citet{naess/etal:2014} we use the notation ${\cal D}^{\rm{XY}}_{\ell} = \ell(\ell+1) C^{\rm XY}_\ell / 2\pi$ where ${\rm XY} \in {{\rm TT,\,TE,\, TB,\, EE,\, EB,\, BB}}$. The covariance matrix for these spectra is estimated using simulations described below, and are compared to an analytic estimate in the Appendix. We then optimally combine the spectra to produce a single 149~GHz power spectrum for each combination XY. The method for doing this coaddition is also described in the Appendix.  The full covariance matrix includes extra terms to account for calibration uncertainty and beam uncertainty.

We `blind' the EB, TB, and BB power spectra throughout our analysis by avoiding estimating them until specific tests are passed.  After testing for internal consistency of the data, described in Section \ref{subsec:null}, we unblind the EB and TB power spectra, and after testing a further suite of null tests described in Section \ref{subsec:nullCust} we unblind the BB power spectra. We do not blind the TT, TE and EE spectra, but do require the same set of consistency and null tests to be passed. We calibrate each power spectrum to \planck\ following \citet{louis/etal/2014}, first cross-correlating the D56 PA2 maps with the \planck\ 143~GHz full-mission intensity maps \citep{planck_mission:2015}, and then by correlating the D56 PA1, D5 and D6 maps with the D56 PA2 map.

\subsection{Simulations}
\label{subsec:sims}

We test the power spectrum pipeline by simulating 840 realizations of the sky. For each one a Gaussian signal is generated on the full sky, drawn from a power spectrum of the sum of the expected signal and foregrounds at 149~GHz. This neglects the non-Gaussian nature of the foregrounds. The D56, D5, and D6 regions are then cut out and projected onto the flat sky, and a Gaussian noise realization added, drawn from the two-dimensional noise power spectrum estimated from the data by differencing different split maps, and weighted by the data hit count maps. These simulations therefore include appropriate levels of non-white noise, but neglect the spatial variation of the two-dimensional power spectrum. Each set of maps is then processed in the same way as the data.

Examining the spectra, we find the dispersion to be consistent with the statistical uncertainty. We construct the data covariance matrix from the simulations, as described further in the Appendix. We also estimate \LCDM\ parameters from 100 of these simulations, to test for parameter bias. Since we only have 149~GHz data here, we fix the residual foreground power to the input value, and vary only the six cosmological parameters. Further, since we use only the ACTPol simulated data, we impose a prior on the optical depth and spectral index, with $\tau=0.08 \pm 0.02$ and $n_s=0.9655 \pm 0.01$\footnote{Note that in our parameter analysis in \S \ref{sec:params} we use an alternative prior for the optical depth, and do not impose a prior on the spectral index.}. We find the parameters are recovered with less than 0.2$\sigma$ bias, where $\sigma$ corresponds to the uncertainties on cosmological parameters for a single simulation. This also tests the validity of our flat-sky approximation. We also extend the parameter set to include the lensing parameter $A_L$, which artificially scales the expected lensing potential as in \citet{calabrese/etal:2008}. We estimate this from each of the simulations, and recover $A_L=1$ to 0.1$\sigma$.

\begin{table} [t!]
\caption{\small Internal consistency tests}
\begin{center}
\begin{tabular}{ll  r  r r }
\hline
\hline
Test & Patch & Spectrum &$\chi^{2}$/dof &P.T.E  \\
\hline
Array      & D56   &TT & 0.90 & 0.69  \\
(PA1-PA2)           &       &EE & 0.74  & 0.91  \\
               &       &TE & 0.64  &  0.98 \\
 &       &TB & 0.89 & 0.69   \\
 &       &EB & 1.40 &  0.03 \\
   &       &BB & 0.60 & 0.99  \\
\\
Array      & D56   &TT & 0.77 & 0.89   \\
(PA1xPA2-PA2)           &       &EE & 0.90  & 0.68  \\
               &       &TE & 1.06 & 0.35  \\
 &       &TB &  0.86 & 0.76  \\
 &       &EB &  1.09 &  0.31 \\
   &       &BB &  0.75 &0.92   \\
\\
\hline
Season       & D56-D5    &TT & 0.88 & 0.71  \\
(S2-S1, PA1)    &       &EE & 1.08 & 0.32   \\
  &       &TE & 0.83 & 0.80   \\
           &       &TB & 0.89 & 0.70   \\
  &       &EB & 1.07 &  0.34 \\
  &       &BB & 0.68 &  0.96 \\
\\
           & D56-D6    &TT & 0.93 &0.62  \\
           &       &EE & 1.09 & 0.30 \\
            &       &TE & 0.98 &  0.51 \\
 &       &TB & 0.94 & 0.60  \\
 &       &EB & 0.96 &  0.56  \\
   &       &BB & 0.99 & 0.49  \\
\hline
\end{tabular}
\end{center}
\label{table:null_nom}
\end{table}

\subsection{Data consistency}
\label{subsec:null}

To identify possible residual systematic effects, we assess the consistency of the power spectra of subsets of our data, splitting the data by array for D56, by season, and by time-ordered-data split.

Splitting the D56 data by array looks for systematic effects that differ between these two arrays, which could include a number of instrumental effects as the two detector arrays were fabricated and assembled independently. 
Here we look at the difference between the PA1 and PA2 power spectra, and compute the covariance matrix of this difference using our simulation suite.
In fact, it was our first analysis of this null test which indicated a difference between the response of the two arrays, and led to our identification of the beam sidelobes (Figure \ref{fig:sidelobe}) that differ between PA1 and PA2. Including the beam sidelobe model we find that this test is passed, as indicated in Table \ref{table:null_nom}.

The season test looks for systematic effects in the array or telescope that vary on long time-scales. The sky coverage is not the same between the two seasons, so to perform the season test we cut out just the part of D56 that overlaps with D5 and D6. The results are reported in Table \ref{table:null_nom} for D56 observed with PA1, and are consistent with null. We also check the difference between the D5 and D6 PA1-S1 spectra and the D56 spectra observed with PA2 in S2, and find no evidence of inconsistency.

\begin{figure}[t!]
	\includegraphics[width=\columnwidth]{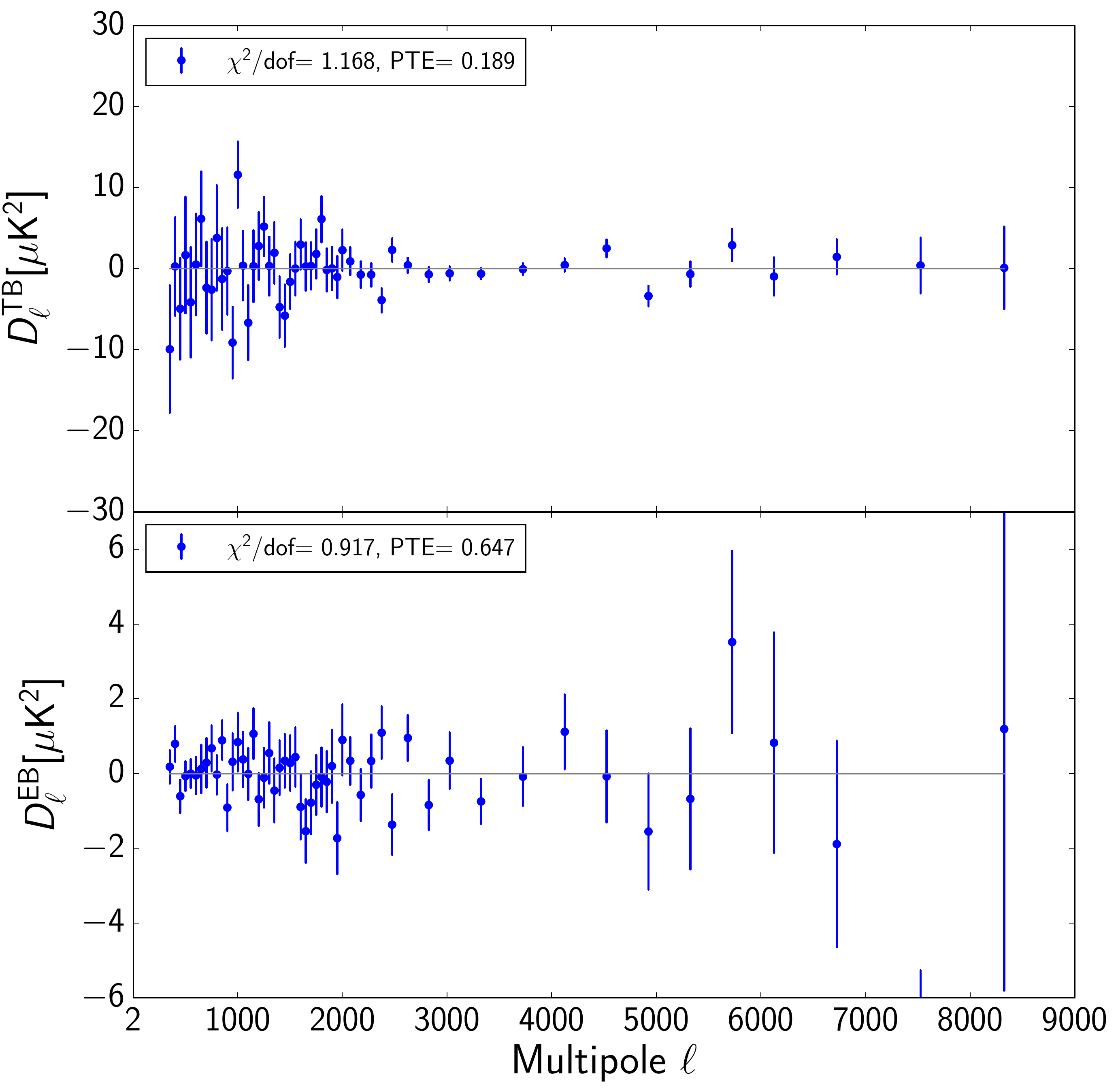}
\caption{The TB (top) and EB (bottom) power spectra, unblinded after internal data consistency checks. The $\chi^{2}$/dof and probabilities to exceed (PTE) are consistent with the null hypothesis for both spectra.}
\label{fig:eb}
\end{figure}

After passing this set of consistency tests, we unblind the EB and TB power spectra, shown in Figure \ref{fig:eb}. The EB and TB power spectra test the polarization angle measurement \citep[e.g.,][]{keating/etal:2013}. This can be biased by Galactic foreground emission, but the effect is estimated to be negligible for ACTPol \citep{abitbol/etal:2016}. We vary an overall offset parameter, and find it to be consistent with zero for all our maps, with $\phi=0.40\pm 0.26^\circ$  for PA1, and $-0.25 \pm 0.36^\circ$ for PA2. We do not re-calibrate the polarization angle, using the original angle estimates as standard. Since these original angle estimates do not yet include a well-characterized systematic uncertainty, we do not estimate cosmological quantities from the EB and TB power spectra.

\subsection{The 149 GHz power spectra}

\label{sec:spectra}
\begin{figure}[ht!]
  \centering
	\includegraphics[width=\columnwidth]{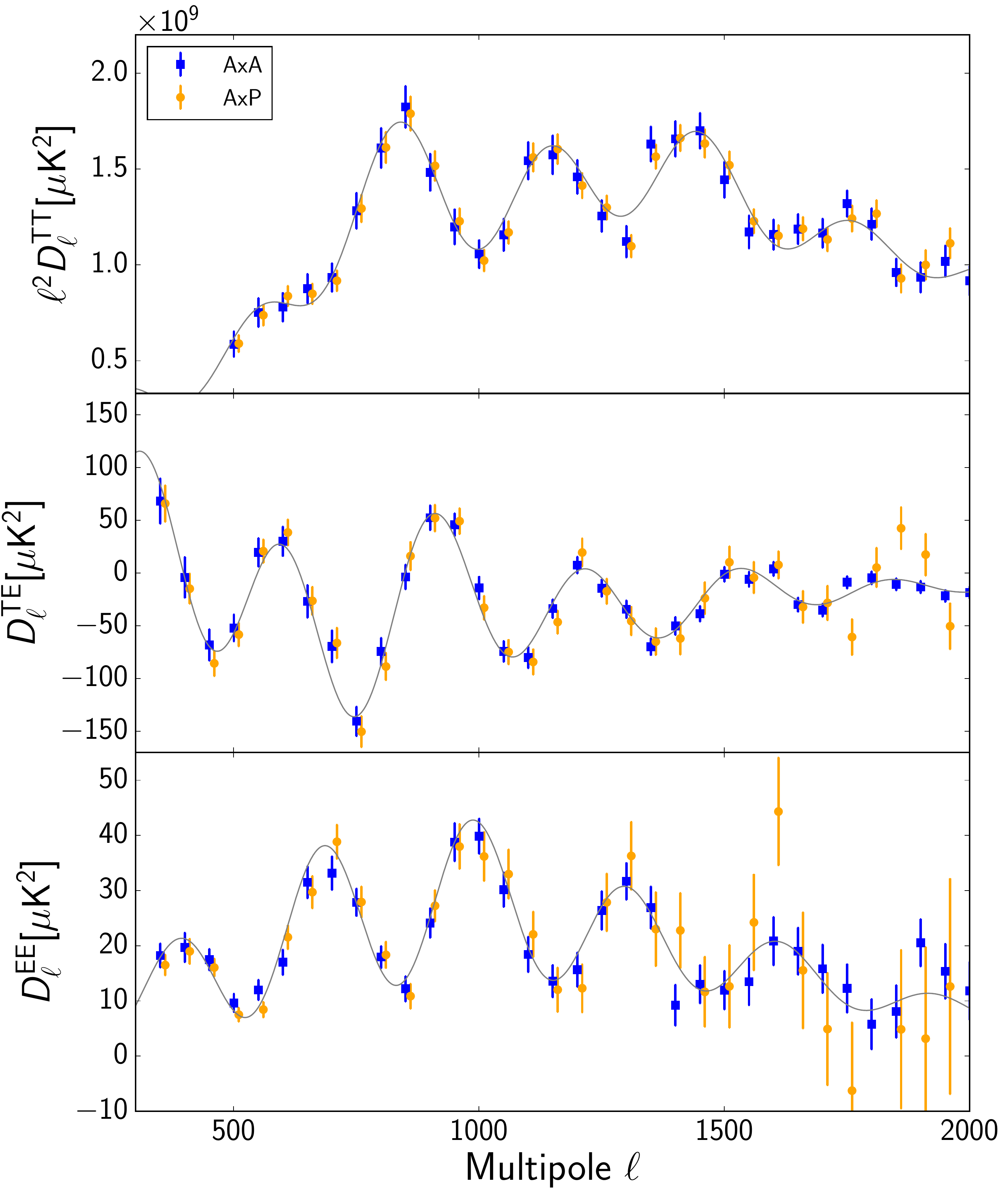}
\caption{Cross-correlation of the D56 PA2 map with the \planck\ 2015 143~GHz temperature and polarization maps. For clarity we shift the ACTxPlanck spectra by $\delta \ell= 10$ compared to the ACTxACT spectra. They are consistent and the relative calibration factor is $0.998 \pm 0.007$ in temperature, defined such that \planck\ is lower than ACT by that factor. }
\label{fig:calib}
\end{figure}

Given the internal consistency of the spectra, we proceed to calibrate the maps by cross-correlating with the Planck-2015  143~GHz temperature maps. The cross correlation of the D56 PA2 maps with the Planck maps is shown in Figure \ref{fig:calib}. Here we follow the same method as in \citet{louis/etal/2014}. We find the ACT x Planck (AxP) cross-spectra to be consistent with the ACT auto-spectra (AxA): their differences have a reduced $\chi^2$ of 0.68, 1.10, 1.17, with PTE of 0.93, 0.31, 0.22,  for TT, TE and EE. No obvious shape dependence or anomalies are detected. The temperature calibration factor is found to be $ 0.998 \pm 0.007$. Cross-correlating the D5, D6, D56 PA1 maps with D56 PA2 gives relative calibrations of $1.002 \pm 0.012$, $0.996 \pm 0.01$,  and $1.009 \pm 0.007$. We then rescale all the maps to have unit calibration.
We do not calibrate our data to \planck\ polarization data, but we test the cross-correlation of the D56 polarization maps with the Planck-2015 143~GHz Q and U maps. The spectra appear consistent, as shown in Figure \ref{fig:calib}, and the correlation implies an ACTPol polarization efficiency of $0.990\pm 0.025$.

\begin{figure}[t!]
	\includegraphics[width=\columnwidth]{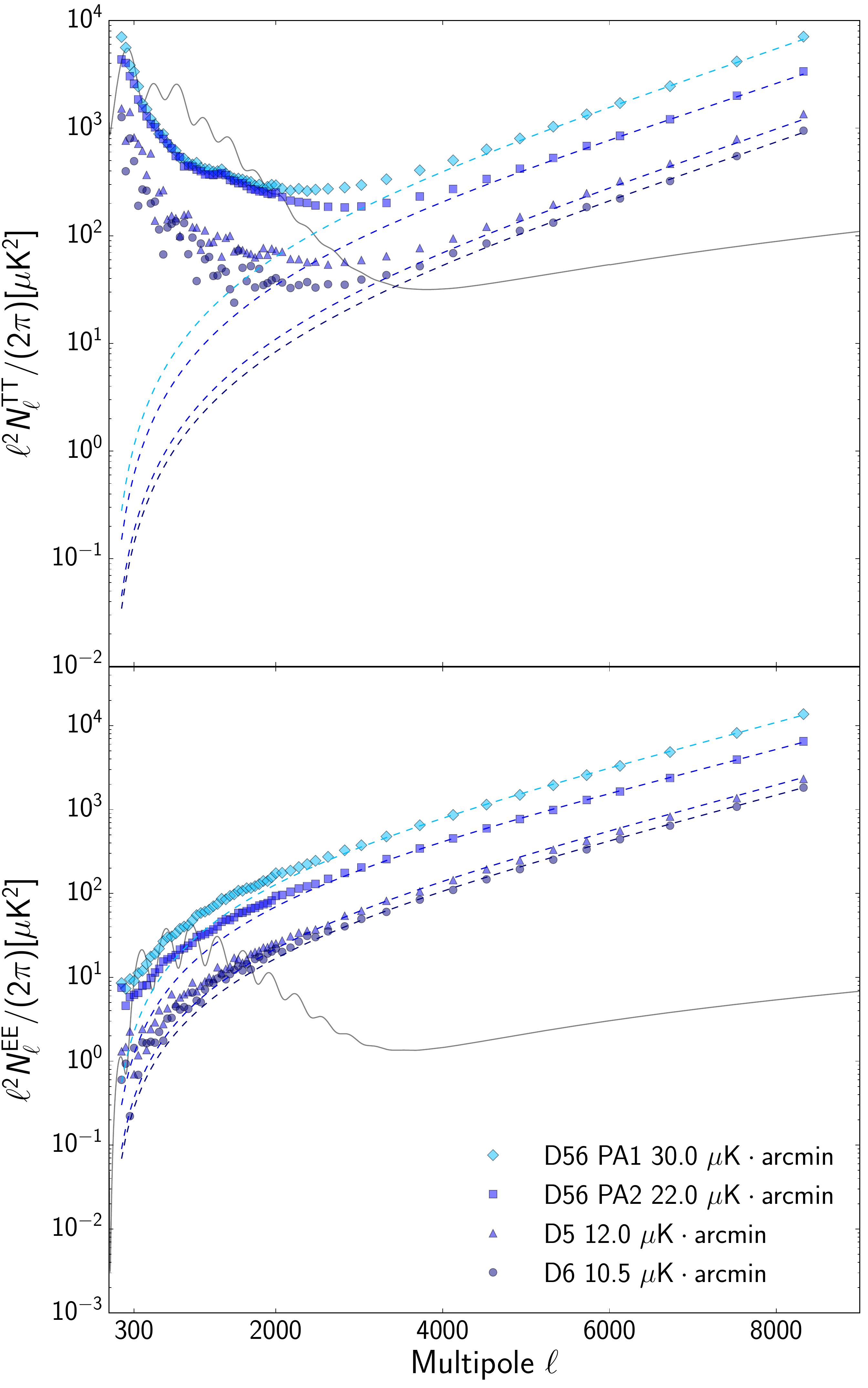}
        \caption{Noise levels in the ACTPol two-season maps, with \LCDM\ theory spectra included for comparison. In temperature the large-scale noise is dominated by atmospheric contamination. In polarization the contamination is significantly lower, and instrumental noise dominates at $\ell \gtsim 1000$. The white noise levels given in the legend are shown with dashed lines.
These noise curves are from the analysis of roughly half the data that passes quality screening procedures from these two seasons. }
\label{fig:noise}
\end{figure}

The noise levels for these maps are shown in Figure \ref{fig:noise}, indicating the dominance of non-white atmospheric noise at scales $\ell<3000$ in temperature. The atmospheric noise is significantly suppressed in polarization, although it dominates the noise power at scales below $\ell\approx 1000$.  A powerful technique for suppressing large scale atmospheric noise contamination in polarization is the
use of a half-wave plate that modulates the polarization at timescales shorter than most atmospheric fluctuations. The Atacama B-Mode Search telescope (ABS) has shown this results in noise power spectra that are white down to large angular scales \citep{2014RScI...85c9901K}. We are currently testing this technique using a subset of ACTPol data taken with a half-wave plate in operation.

\begin{figure*}[htp!]
\centering
	\includegraphics[width=0.98\textwidth]{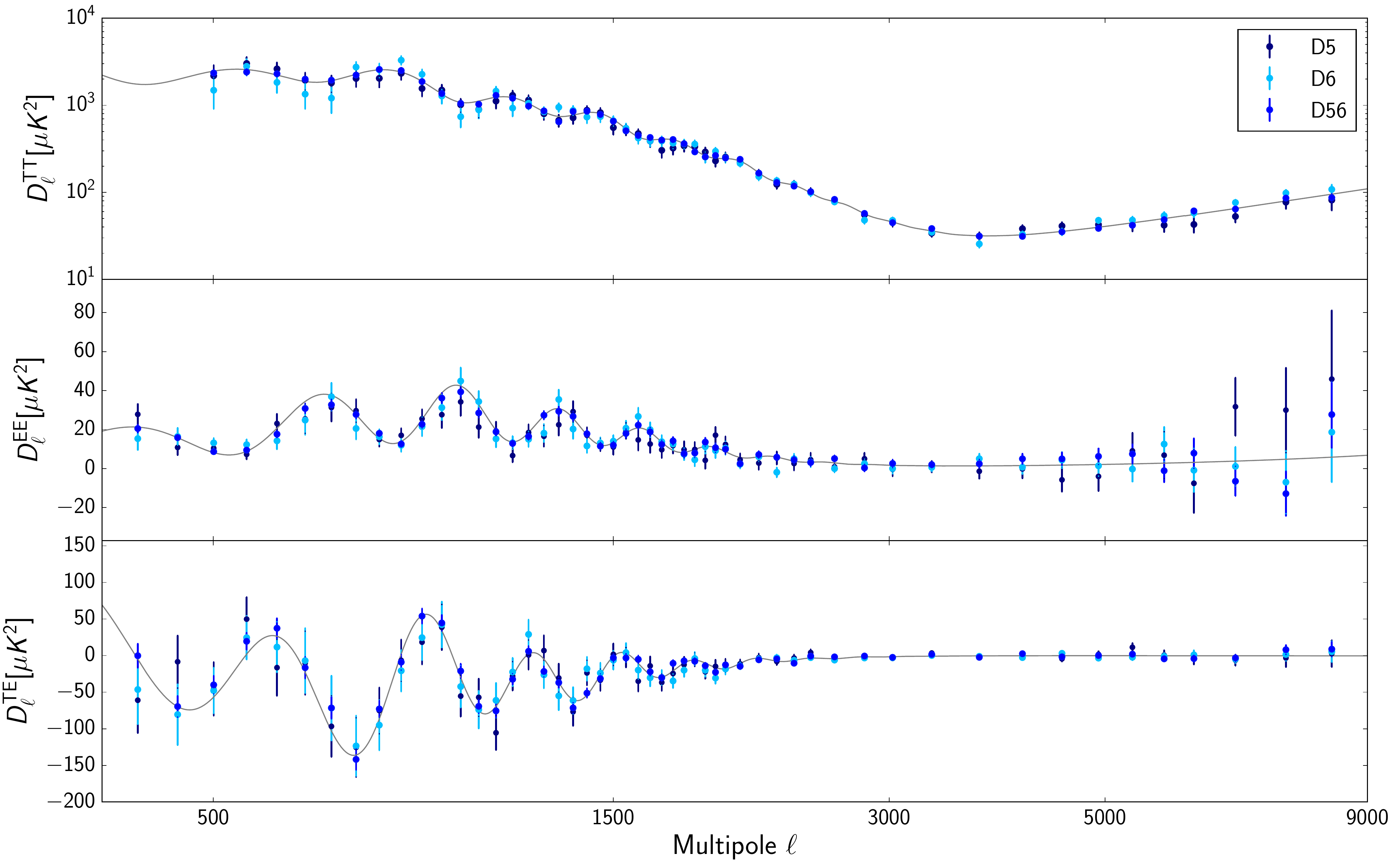}
   \caption{The ACTPol power spectra (TT,TE,EE) for individual  D56, D5, and D6 patches. For D56 the PA1 and PA2 data have been co-added. The solid lines correspond to the ACTPol best-fit \LCDM\ model, including the foreground contribution at 149~GHz.}
\label{fig:specpatch}
\end{figure*}

The TT, TE, and EE power spectra for the calibrated ACTPol maps in each region are shown in Figure \ref{fig:specpatch}, corrected for the transfer function. The temperature and polarization acoustic peak structure is clearly seen in all the maps, with six acoustic peaks measured in polarization. As expected, the D56 maps provide the best estimate of the power at large scales, due to the larger sky area. At smaller scales the deeper D5 and D6 maps contribute more statistical weight. The reference model shown is the best-fitting \LCDM\ model with best-fitting foreground contribution, described in section \ref{sec:like}.

\begin{figure*}[htbp!]
\centering
	\includegraphics[width=0.98\textwidth]{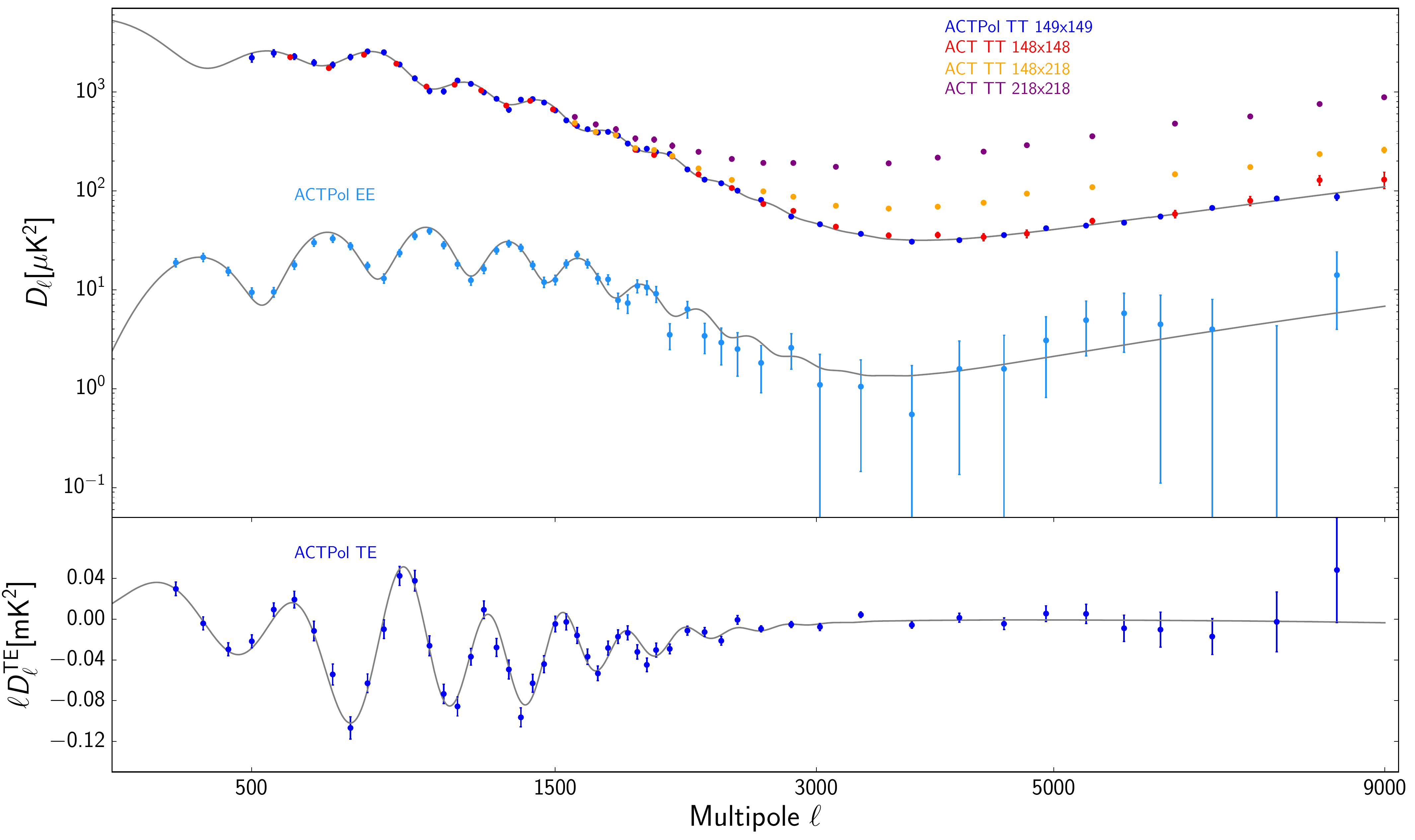}
   \caption{Two-season optimally combined 149~GHz power spectra for temperature and E-mode polarization (top), and TE cross-correlation (bottom). The solid lines show the ACTPol best-fit \LCDM\ model including the 149~GHz foreground model. The best-fitting foreground model for the 218~GHz data is not shown.}
\label{fig:coadd}
\end{figure*}

The optimally combined spectra are shown in Figure \ref{fig:coadd} for temperature, E-mode polarization, and the TE cross-spectrum, and reported in Table \ref{tab:spectra}.  Here, the temperature data have the expected residual foreground contribution that dominates at scales smaller than $\ell \sim 3000$. For comparison, the ACT MBAC temperature data are also shown for the coadded ACT-Equatorial and ACT-South spectra, including 220~GHz data \citep{das/etal/2014}.

\subsection{Real-space correlation}

\begin{figure}[t!]
\includegraphics[width=\columnwidth]{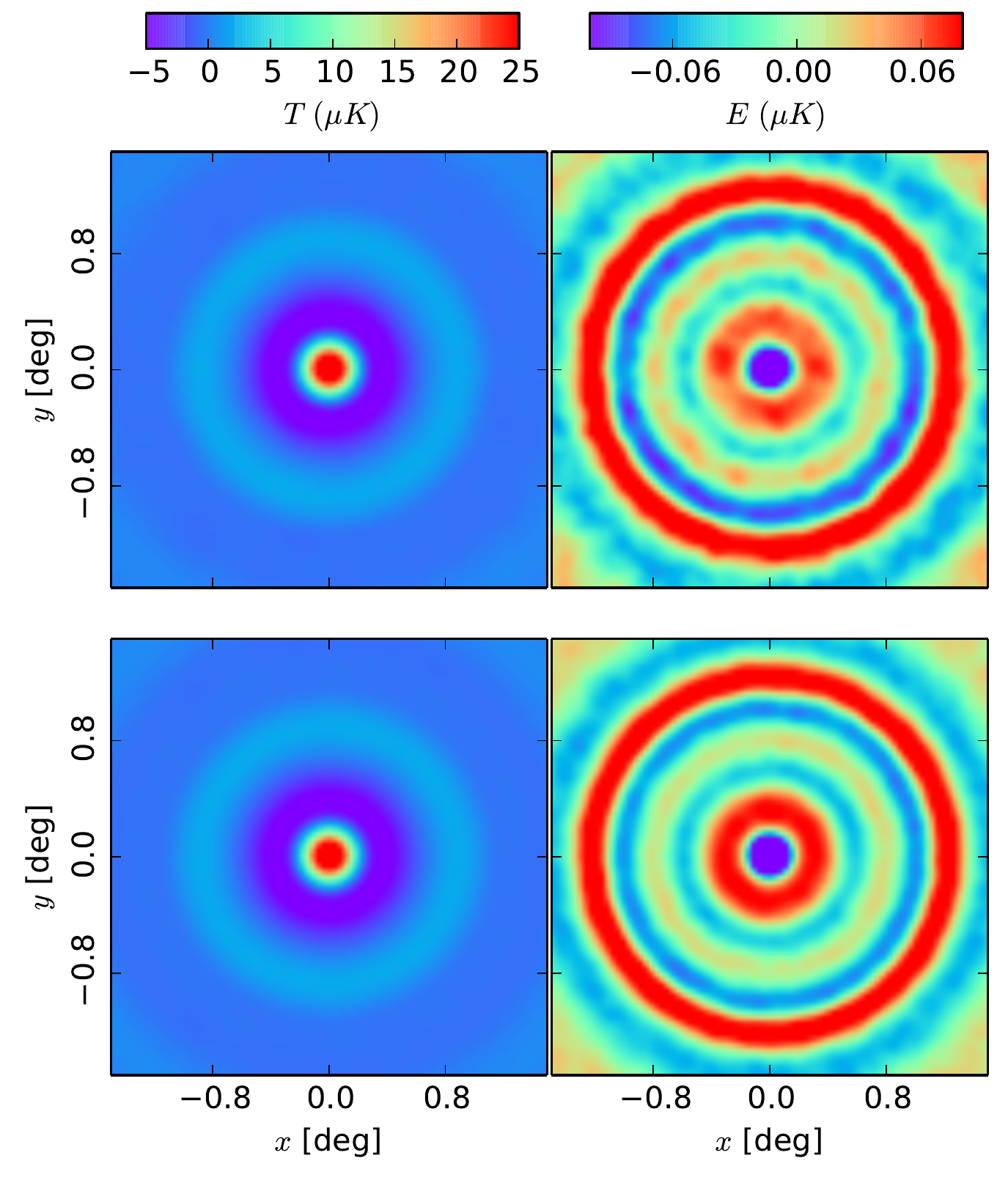}
 \caption{The temperature, $T$, and $E$-mode polarization maps stacked around randomly selected field points in the temperature map. The sign of the map is reversed when it is stacked around a cold field point with $T < 0$. These provide direct estimates of the $TT$ and $TE$ correlation functions. The top panels show the result from the coadded D56 PA1 and PA2 maps smoothed with a FWHM $5$ arcmin Gaussian beam and high-pass filtered with $\ell_{\min}=350$. The bottom panels show the average of 30 simulations generated with Planck-2015 $\Lambda$CDM parameters, with noise simulations estimated from the ACTPol data.}
\label{fig:stack}
\end{figure}

The \wmap\ team first stacked temperature and polarization data on temperature hot and cold spots to help visualize acoustic patterns in the data~\citep{komatsu/etal:2009}. With \planck\ data, the noise of the stacked 2D images was considerably reduced~\citep{2016A&A...594A..16P}. We now repeat this exercise with the ACTPol data. Although such patterns do emerge in the ACTPol data, there are not as many extrema to stack on and the result is noisier than for \planck. To decrease the noise, and provide a direct measure of the $TT$ and $TE$ cross correlation functions $C^{TT}(\mathbf{\theta})$ and $C^{TE}(\mathbf{\theta})$, we instead stack on a much larger set, using randomly chosen temperature field points.

Figure~\ref{fig:stack} shows the D56 temperature and $E$-mode polarization maps stacked on a uniformly chosen sample of `hot' points with $T>0$, and, with flipped sign, on a `cold' sample with $T < 0$. For $E$-polarization, with enough points the result should converge to the ensemble average given the $\{ T \}$ constraints, $ \langle E (\mathbf{\theta}) |\{ |T| \} \rangle = C^{TE}(\mathbf{\theta}) \langle |T| \rangle /C^{TT}(0)$, where $\langle |T| \rangle$ is the ensemble average of $|T|$ at randomly chosen field points ($\sqrt{2/\pi}\, C^{TT}(0)^{1/2}$). A similar result holds for the mean temperature. Around each stack-point, the $T$ and $E$ fields are randomly rotated, and so should be spherical, as they clearly are.

The rings in the patterns depend upon the low-pass and high-pass filtering of the maps, but reflect the acoustic patterns in a more direct way than stacking on extrema. To demonstrate that our ACTPol stacks agree with theoretical expectations, in the lower panels of Fig~\ref{fig:stack} we compare an average of 30 $\Lambda$CDM simulations processed in the same way, with ACTPol noise estimated from map differences included. By angle-averaging at each radius we generate direct isotropic correlation function estimates in excellent agreement with the simulations. By varying temperature thresholds, rotation strategies, map selections and data cuts, the stacked maps help show the robustness of the ACTPol data sets. Note that we do not yet stack E on E field points because of the higher noise levels.

\subsection{Galactic foreground estimation}
\label{subsec:dust}

\begin{figure}
	\includegraphics[width=\columnwidth]{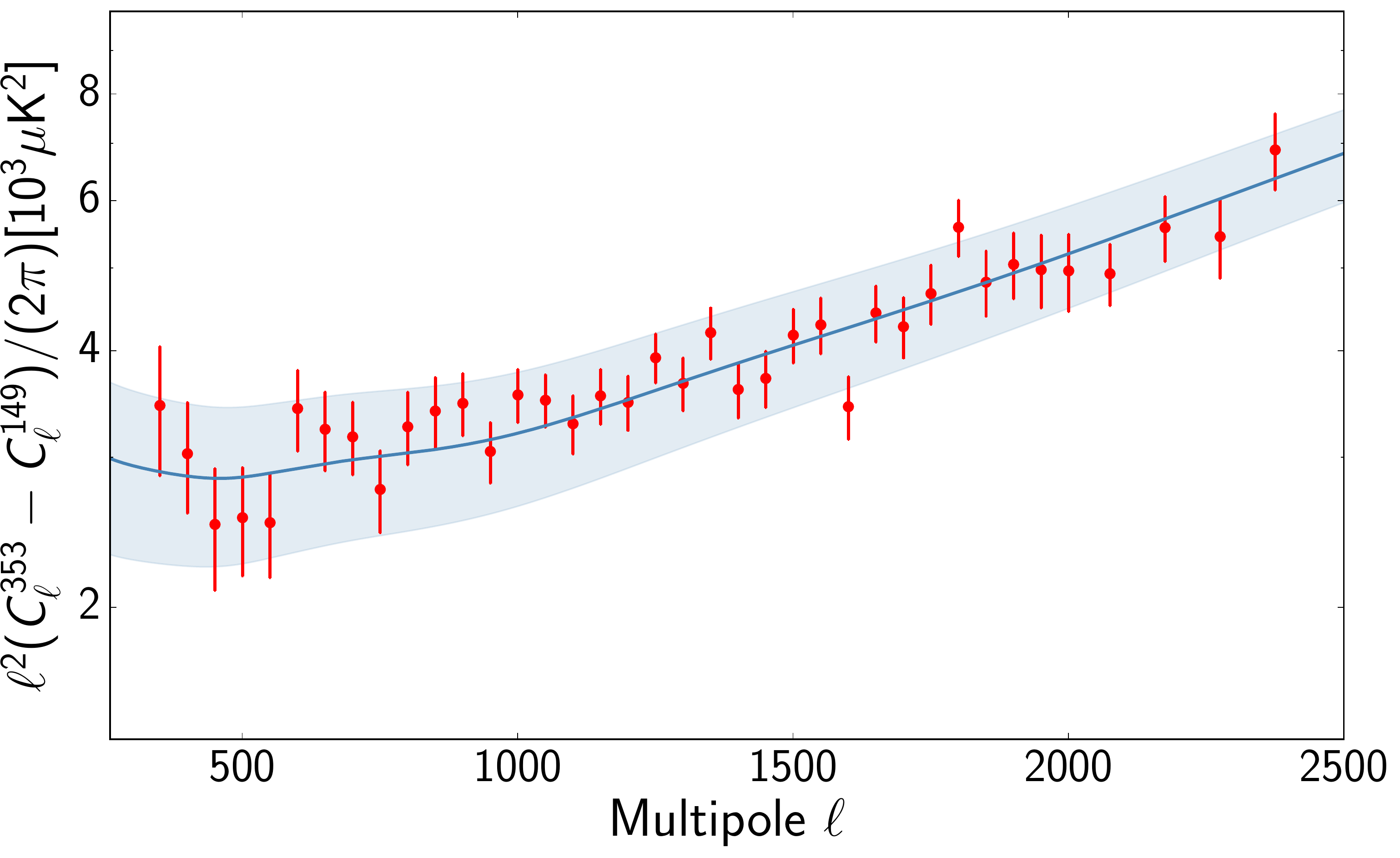}
\caption{Difference between the Planck 353 GHz and ACTPol 149 GHz power spectra in the D56 patch. The band shows the dust+CIB foreground model used for the \cite{das/etal/2014} ACT analysis, with the CIB clustered component template replaced to match that used in the \planck\ analysis. The width of the band reflects the 1$\sigma$ uncertainties on the parameters of the model. We find good agreement between this model and the data.}
\label{fig:dust}
\end{figure}

We estimate the level of thermal dust contamination in the power spectrum using the \planck\ 353~GHz dust maps \citep{planck_mission:2015}. We compute the difference between the power spectrum of the Planck 353~GHz maps and the ACTPol power spectrum at 149 GHz, following a similar method to \citet{louis/etal/2013}. 
The result is shown in Figure \ref{fig:dust}. The difference between the two power spectra is dominated by CIB fluctuation and Galactic cirrus emissions at 353 GHz. On large and intermediate scales, the contributions from other signals are subdominant and can be neglected. The shaded band represents the CIB and dust model from \citet{dunkley/etal/2013}, valid for the overlapping ACT-Equatorial region, with the exception of the CIB clustered source template that we have replaced to match the one used in the nominal \planck\ analysis \citep{2014A&A...571A..30P}. We find this model to be a good fit to the 353-149 differenced spectrum, so use the same ACT-Equatorial dust level as a prior in the likelihood. In E-mode polarization, we find that the dust signal is negligible for all scales of interest.

\subsection{Null tests}
\label{subsec:nullCust}

\begin{table}[t!]
\caption{\small Null tests using custom maps (PA1, PA2)}
\begin{center}
\begin{tabular}{ll  r r r r}
\hline
\hline
Test & Spectrum & PA1& & PA2&  \\
 &  &$\chi^{2}$/dof &P.T.E &$\chi^{2}$/dof &P.T.E  \\
\hline
Scan pattern 1 
          &TT & 0.82  &  0.83  & 1.00 & 0.47\\
 v Scan pattern 2:                    &EE &0.91  &0.66  & 0.72 & 0.94  \\
 (0-1)x(2-3)                 &TE & 0.99  & 0.49  &0.80  & 0.85\\
 &       TB & 1.13  &  0.25 &0.86 &0.76\\
 &       EB &  1.15 &  0.21 &0.93 &0.61\\
   &       BB &  0.66 &  0.97 &0.83 &0.81\\
Scan pattern 1            &TT &1.13  &  0.24 &1.19 & 0.17\\
v Scan pattern 2:                    &EE &0.67  & 0.97 &1.12 &0.25 \\
 (0-3)x(1-2)                 &TE & 0.99  &  0.50 &0.83 & 0.80\\
 &       TB & 0.85  & 0.77  &0.81 & 0.84\\
 &       EB &  0.95 &   0.58 & 0.98 & 0.53\\
   &       BB &  0.96 & 0.55 & 0.75 & 0.91 \\
 \\
\hline
Detectors: 
          &TT &0.98  & 0.51  & 0.89 & 0.69\\
 Fast v slow                    &EE & 0.78&0.88  &0.72 & 0.94  \\
           &       TE & 0.94  &  0.59 &0.87 & 0.74\\
 &       TB & 1.07 & 0.34  & 0.78 & 0.88\\
 &       EB & 0.81 &  0.84 & 0.68 & 0.96 \\
   &       BB &  1.02 & 0.42  &1.00 & 0.48 \\
\\ 
\hline
PWV: 
           & TT & 0.99 & 0.49 & 1.18 & 0.18 \\
 High v low           &       EE & 0.84 & 0.78 &0.90 &0.68 \\
           &       TE & 0.72  & 0.94 &0.71 & 0.94 \\
 &       TB &  0.75&  0.91 &0.77 &0.89\\
 &       EB &  0.98 &  0.52 & 0.96 & 0.56\\
   &       BB & 0.65 &  0.98 &0.94 & 0.60 \\
\\
\hline
Pick up: 
             &TT &1.14 & 0.22  &0.94 & 0.61\\
         &       EE &  0.83& 0.81  &0.64 & 0.98 \\
                  &TE &0.87  &  0.74  &0.88 & 0.72\\
 &       TB &0.83  & 0.80 &0.95 & 0.58 \\
 &       EB & 0.64  &0.98  &0.95 & 0.58 \\
   &       BB & 1.00  &0.47  &0.83 & 0.81 \\
\\
\hline
Moon: 
             &TT & 0.82 &  0.82 & 1.08 &0.32\\
more aggressive            &       EE & 1.40 & 0.03 &1.18 & 0.17 \\
cut                  &TE &1.30  &   0.07 &0.68 &0.97 \\
 &       TB &0.92  &  0.64  &0.91 & 0.66\\
 &       EB & 1.01  & 0.45  &0.96 &0.55\\
   &       BB & 0.90  & 0.67  &1.22 &0.13\\
\\
\hline

Wafers: 
          &TT & &&1.02 &  0.44   \\
 Hex1+hex3                   &EE &&& 1.08 &  0.33 \\
   v hex 2+semis                 &TE &&& 1.29 & 0.07  \\
 &       TB &&& 0.59  & 0.99  \\
 &       EB &&&  1.03 & 0.42  \\
   &       BB &&&  0.54 &  0.99  \\
\\
\hline
\end{tabular}
\end{center}
\label{table:null_new_pa1}
\end{table}

\begin{figure}[ht!]
	\includegraphics[width=\columnwidth]{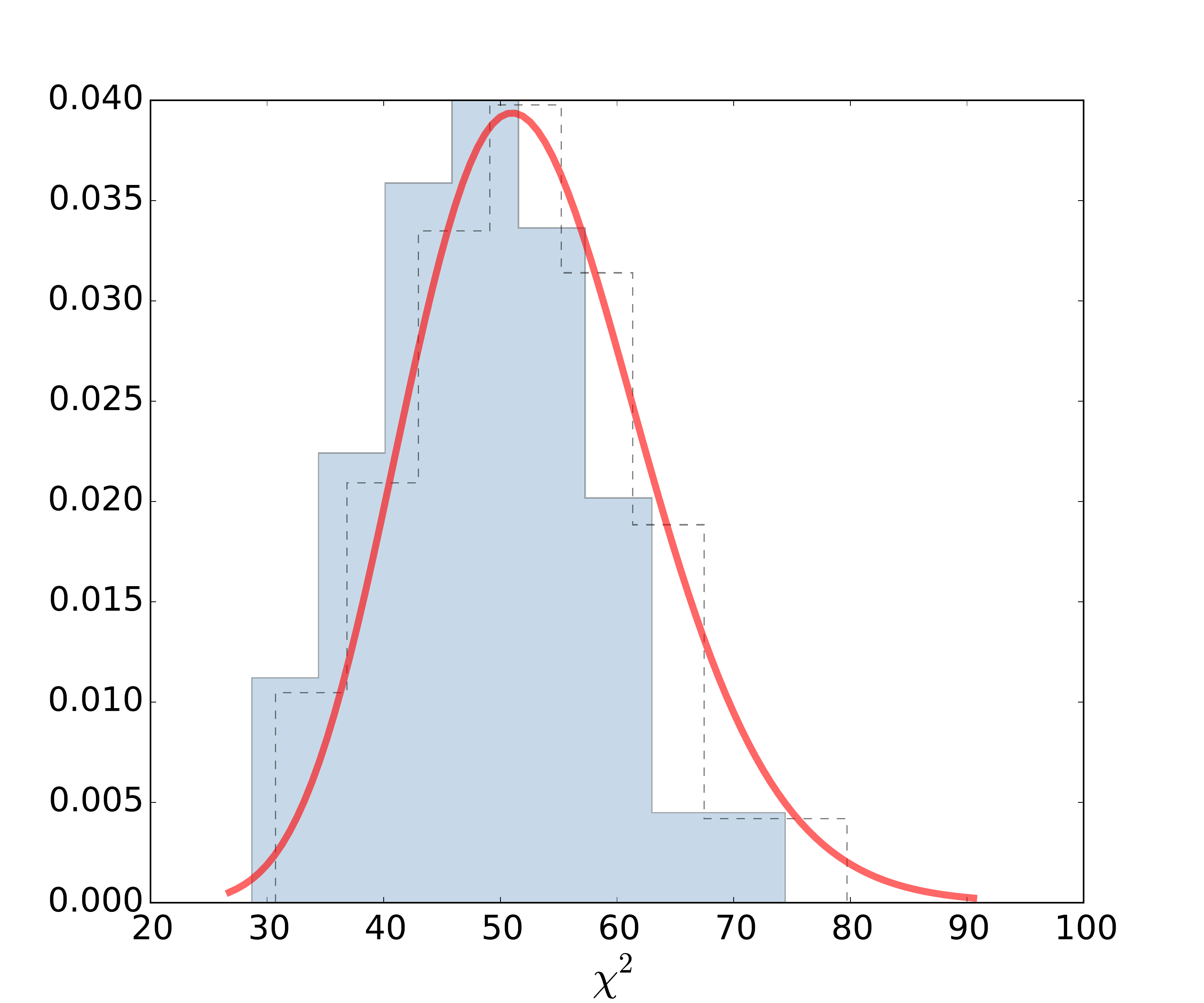}
   \caption{Distribution of the $\chi^{2}$ for the null tests described in Section \ref{subsec:nullCust}. The smooth line represents the expected distribution if the null tests were uncorrelated. The dashed black histogram shows our null test distribution after rescaling the errors by $3\%$. We interpret this as an estimate of the uncertainty on our errors.\\}
\label{fig:chi2}
\end{figure}

We make an additional suite of maps to identify further possible systematic contamination. The first set of tests splits the data into two parts. We test for dependence on the scan pattern by splitting the data for D56 into the two different elevations. We then test the effect of detector performance, making maps from detectors with faster and slower time constants. The threshold is chosen to give roughly equal statistical weight to each subset, splitting at 80 Hz. We test the impact of weather and atmospheric noise on the data by splitting on precipitable water vapor level (PWV). We choose a threshold of 0.8~mm, again to give equal weight to both halves.

We test the impact of internal telescope pick-up fluctuations by splitting each array into two groups of detectors based on their qualitative behavior.  We also run an additional null test for PA2, testing the different detector wafers by splitting the data based on their thermal conductivity to the bath. (For this specific test, the number of detectors in PA1 is too small to pass the internal cuts of the map-maker.)
Finally we test the effect of applying a more aggressive moon cut. In all these cases we generate four splits for each map subset, so the power spectrum is estimated from four splits as usual. 
The $\chi^{2}/\textrm{dof}$ and PTE of all these null tests are reported in Table \ref{table:null_new_pa1}.

We do not find any indication of contamination from any of these systematic effects in the power spectrum. The $\chi^{2}$ distribution for this set of null tests is shown in Figure \ref{fig:chi2}. The distribution is close to expectation, but we find that  the measured and predicted $\chi^{2}$ distribution fit best if we reduce the error bars by $\approx 3\%$. We interpret this
as an estimate of the uncertainty on our errors.

\subsection{Effect of aberration}

\begin{figure}
	\includegraphics[width=\columnwidth]{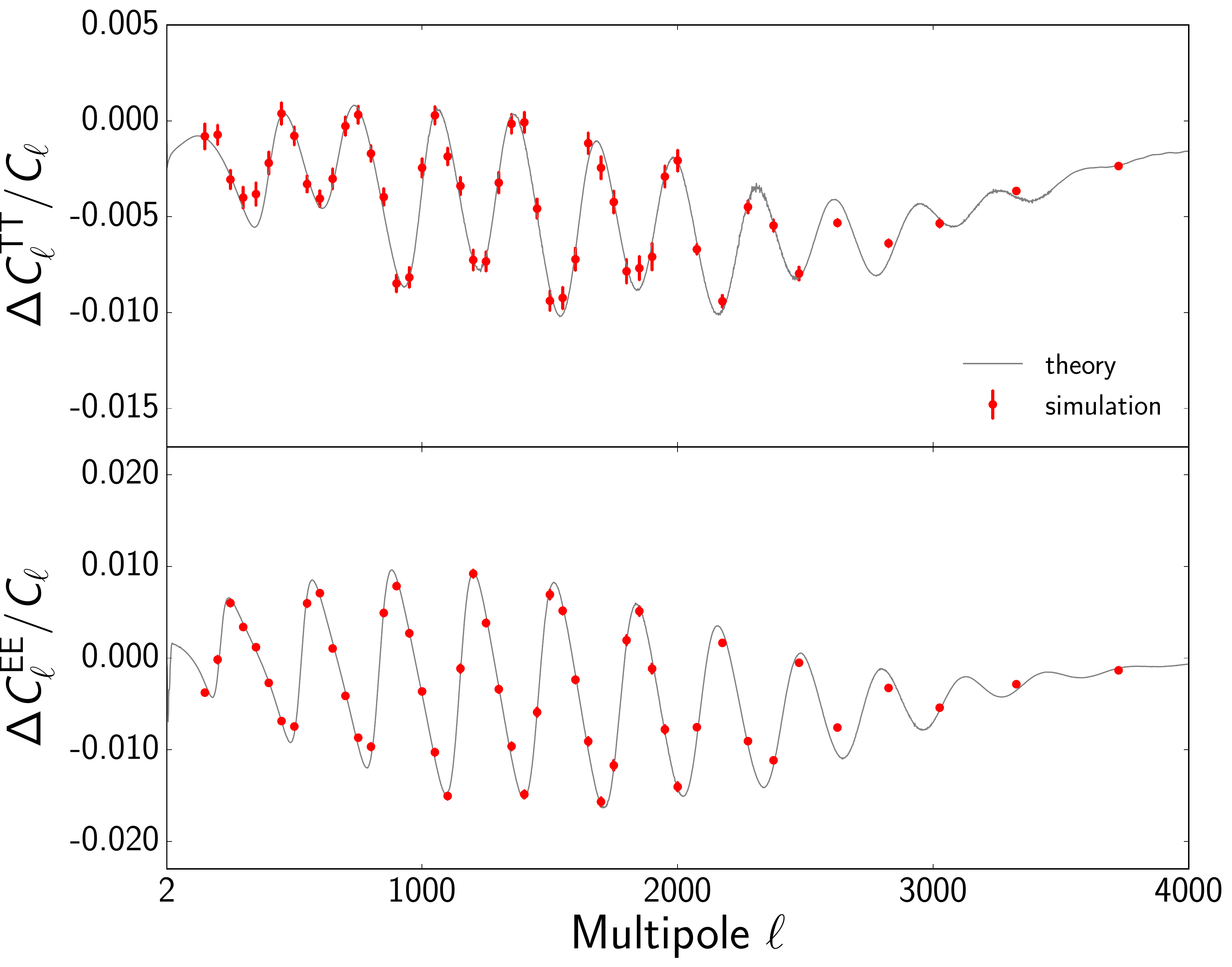}
\caption{Effect of aberration on the TT and EE CMB power spectra due to our proper motion with respect to the CMB. Our aberrated simulations agree with the analytical estimate of the expected effect.}
\label{fig:aberration}
\end{figure}

The observed power spectra are affected by aberration due to our proper motion with respect to the CMB last scattering surface. We move at a speed of 369 km/s  along the direction $\bm{d} = (l,b) = (264^{\circ},48^{\circ})$   (e.g. \cite{2014A&A...571A..27P}).
  This motion induces a kinematic dipole of the form $\cos\theta =(\bm{d} \cdot \bm{n})$, where $\bm{n}$ is the vector position of each pixel. Aberration results in an angle-dependent rescaling of the multipole moments $\ell$ and its effect on the power spectrum can be approximated as 
\ba
\frac{\Delta C_{\ell}}{C_{\ell}}= - \frac{ d \ln C_{\ell}}{d \ln \ell} \beta \langle \cos\theta \rangle
\ea
\citep{2014PhRvD..89b3003J}, where $\beta=v/c$ and $ \langle \cos\theta \rangle= -0.82$ in D56, $-0.97$ in D5 and $-0.65$ in D6, where the average is taken over the solid angle of each ACTPol patch.
We generate a set of 120 aberrated simulations, compute their power spectra and compare it to the power spectra of non-aberrated maps. The result is presented in Figure \ref{fig:aberration} together with the analytical estimate. We use this set of simulations to correct our power spectra for the aberration effect, such that $\hat{C}_{\ell}= C_{\ell}-\Delta C_{\ell}$. In earlier releases the effect was negligible and we did not correct for it. Section \ref{subsec:lcdm} discusses the impact of this correction on cosmological parameters.

\subsection{Unblinded BB spectra}
\label{subsec:bb}

\begin{figure}
	\includegraphics[width=\columnwidth]{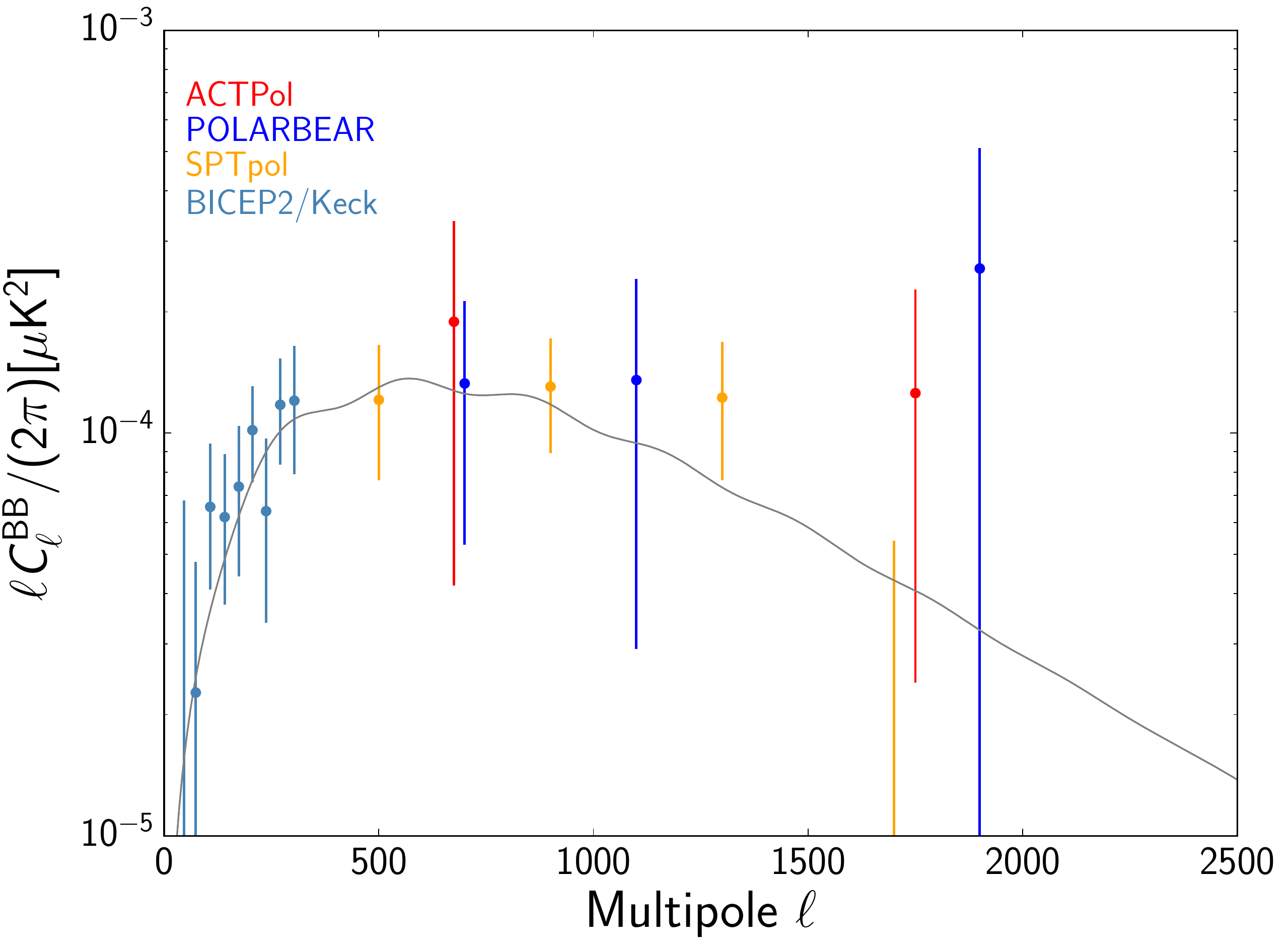}
        \caption{ Unblinded ACTPol BB power spectra compared to measurements from {\textsc POLARBEAR} \citep{pbear-eebb/2014}, SPTpol \citep{keisler/etal:2015} and {\textsc {BICEP2}}/Keck array \citep{2016PhRvL.116c1302B}. The solid line is the \planck\ best fit \LCDM\ model. The ACTPol data are consistent with expectation and deviate from zero at $2 \sigma$.}
\label{fig:bb}
\end{figure}

We unblind the B mode power spectrum at the end of the analysis. The spectrum is shown in Figure~\ref{fig:bb} along with B mode measurements from \citet{pbear-eebb/2014}, SPTpol \citep{keisler/etal:2015} and {\textsc BICEP2}/Keck array \citep{2016PhRvL.116c1302B}. We fit for an amplitude in the multipole range $500<\ell<2500$, where Galactic and extragalactic contamination is minimal, using the lensed B mode \LCDM\ prediction. We find $A =2.03 \pm  1.01$. This amplitude is consistent with expectation, but the significance of the fit is not high enough to be interpreted as a detection.

%% file: likelihood.tex
\section{Likelihood}
\label{sec:like}

\begin{figure*}[ht!]
\centering
	\includegraphics[width=0.87\textwidth]{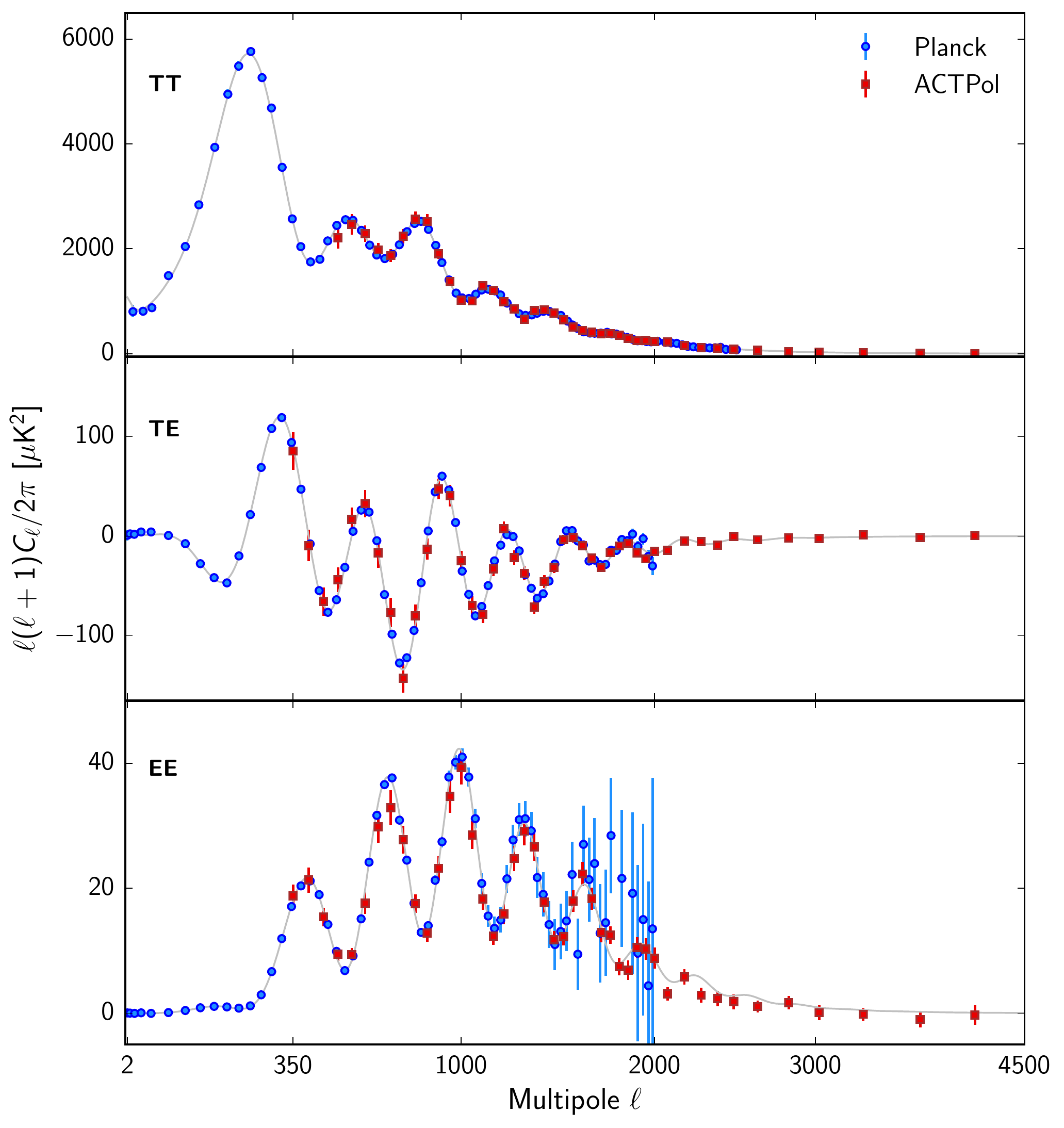}
\caption{Comparison of ACT CMB power spectra (combining MBAC and ACTPol data) with \planck\ power spectra. The uncertainties for ACT are lower than for \planck\ at scales $\ell \gtsim 1500$ in polarization. The theory model is the \planck\ 2015 TT+lowTEB best-fit \citep{planck_cosmo:2015}. The small-scale power spectra have also been measured by SPTpol \citep{2015ApJ...805...36C}.}
\label{fig:planck}
\end{figure*}

We first construct a likelihood function to describe the CMB and foreground emission present in the 149~GHz power spectrum. To improve the estimation of the CMB part, we then add intensity power spectra estimated at both 150 and 220~GHz by the previous ACT receiver, MBAC.

Using these multi-frequency data we estimate the foreground-marginalized CMB power spectrum in TT, TE, EE for ACT, for both the MBAC and ACTPol data. We then combine this likelihood with the data from \wmap\ and \planck. 

\subsection{Likelihood function for 149~GHz ACTPol data}
\label{subsec:actpol}

Following \citet{dunkley/etal/2013}, we approximate the 149~GHz likelihood function $L=p(d|C^{\rm th}_\ell)$ as a Gaussian distribution, with covariance described in Sec.~\ref{sec:spectrum}. We neglect the effects of variation in cosmic variance among theoretical models. The likelihood for the data given some model spectra $C_\ell^{\rm th}$ is given by 
\be
 -2 \ln {L} = (C_b^{\rm th} - C_b)^{\rm T} 
 \mat{\Sigma}^{-1}(C_b^{\rm th} -
  C_b) + \ln \det\mat{\Sigma},
\label{eqn:like}
\ee
where the bandpower theoretical spectra are computed using the bandpower window functions $w_{b\ell}$,  $C_b^{\rm th} = w_{b\ell}C_\ell^{\rm th}$ , as in \citet{das/etal/2014}. We include a calibration parameter $y$ that scales the estimated data power spectra as $C_b \rightarrow y^2{C}_b$ and the elements of the bandpower covariance matrix as $\mat{\Sigma}_{bb} \rightarrow y^4\mat{\Sigma}_{bb}$.  We impose a Gaussian prior on $y$ of $1.00\pm0.01$, using the estimated error from the calibration of ACTPol to \planck.

Since we will include data from MBAC data at 150~GHz and 220~GHz, we write the model spectrum as the sum of CMB and foreground terms, following the approach in \citet{dunkley/etal/2013}. We use the same intensity foreground model that includes Poisson radio sources, clustered and Poisson infrared sources, kinetic and thermal Sunyaev Zel'dovich effects, and Galactic dust. This model has six free extragalactic foreground parameters: an amplitude for each of tSZ and kSZ spectra, an amplitude for each of the Poisson and clustered infrared spectra, an emissivity index for the infrared sources, and a cross-correlation coefficient between the tSZ and clustered infrared emission. The amplitude for the radio source spectrum is also varied with a prior based on observed source counts, and the spectral index is held fixed. The Galactic dust intensity level has a parameter for each different region (the ACTPol D56 region and the two MBAC ACT regions known as ACT-South and ACT-Equatorial), varied with a prior based on the higher frequency observations. This model all follows \citet{dunkley/etal/2013}.

We extend the model to include polarization foregrounds relevant for the ACTPol data, including a single Poisson source term as in \citet{naess/etal:2014} in EE. We allow for an additional Poisson source term in TE that can take both positive and negative values, although this contribution is expected to be negligible. 

A similar approach was used for the \planck\ analysis \citep{planck_cosmo:2015}, which also included ACT and SPT data, but the foreground model we use for ACT differs in the following few ways. Following \cite{dunkley/etal/2013} we use an alternative cosmic infrared background clustering template that differs at large scales, and an alternative thermal SZ template from \cite{battaglia/etal:2010}. This is, however, similar in shape to the \citet{efstathiou/migliaccio:2012} template used in the \planck\ analaysis. As in \citet{dunkley/etal/2013} we also describe the Poisson source components by using an amplitude and a spectral index for each of the radio and infrared components, rather than a free Poisson amplitude at each frequency and cross-frequency as done for \planck.

\subsection{CMB estimation for ACTPol data}
\label{subsec:actpol}

We combine the data from ACTPol and MBAC in the D56 region to estimate simultaneously the CMB bandpower and the foreground parameters, following \citet{dunkley/etal/2013} and \citet{calabrese/etal/2013}.

We write the likelihood as 
\be
-2 \ln L = -2 \ln L({\rm ACTPol}) - 2 \ln L ({\rm MBAC}).
\ee
Here the MBAC data includes both the ACT-S and ACT-E data at 150 and 220 GHz, and the 150-220 GHz cross-correlation. 

We use the Gibbs sampling method of \citet{dunkley/etal/2013} to simultaneously estimate the CMB bandpowers and the foreground parameters. We marginalize over the foregrounds to estimate the CMB bandpowers and their covariance matrix. We measure the EE Poisson power to have $A_p=1.10\pm0.34$, defined in units of $\mu$K$^2$ for $D_{3000}$. This is evidence for Poisson power in the case where no polarized sources are masked. In the analysis of SPTpol data in \citet{2015ApJ...805...36C}, sources with unpolarized flux brighter than 50 mJy are masked at 150 GHz, and an upper limit of $A_p<0.4$ at 95\% CL was found. For ACTPol we find the TE power to be consistent with zero, with $A_{TE}=-0.08\pm 0.22$ at the same $\ell=3000$ scale.

The marginalized spectra are shown in Figure~\ref{fig:planck} and reported in Table~\ref{tab:cmb_spectra}. Figure~\ref{fig:planck} also shows how the ACTPol data compare to \planck\ TT, TE and EE data. Due to its larger sky coverage the \planck\ uncertainties are smaller at large scales, but at scales $\ell \gtsim 1500$ the ACTPol uncertainties in polarization are smaller.

\subsection{Foreground-marginalized ACTPol likelihood}
\label{subsec:foreg}

Following \citet{dunkley/etal/2013}, we use the marginalized ACTPol spectrum to construct a new Gaussian likelihood function. The only nuisance parameters in this likelihood are an overall calibration parameter,  and a varying polarization efficiency parameter. 
The likelihood includes data in the angular range $350<\ell<4000$, using scales where the distribution of the marginalized spectra is Gaussian to good approximation.

\subsection{Combination with \planck\ and \wmap}
\label{subsec:planck}

For some investigations we combine the ACTPol data with \wmap\ and \planck\ data. This is done by adding the log-likelihoods, since there is little overlap in angular range and since the ACTPol survey area represents a small fraction of the sky observed by \planck. We use the \planck\ temperature data \citep{planck_like:2015} at $2<\ell<1000$ as a baseline, and over the full range $2<\ell<2500$ for other combined-data tests. We label \planck\ temperature at $2<\ell<1000$ `PTTlow'. We use the public CMB-marginalized `plik-lite' likelihood, constructed using our same marginalization method.
The CMB likelihood is then
\ba
-2 \ln L &=& -2 \ln L({\rm ACTPol}) \nonumber\\
&&- 2 \ln L ({\rm PlanckTT_{2<\ell<1000,2500}}) \,.
\ea

For TE-only tests we use the \wmap\ likelihood at $\ell<800$, since it includes TE cross-correlation data \citep{hinshaw/etal:2013}.

Instead of explicitly using the large-scale TE and EE polarization data from \planck\ or \wmap\  we choose to impose a prior on the optical depth of $\tau=0.06\pm0.01$, derived from the \planck-HFI polarization measurements \citep{planck_reion:2016}.

%% file: params.tex
\section{Cosmological parameters}
\label{sec:params}

We use standard MCMC methods to estimate cosmological parameters, using the CosmoMC numerical code \citep{lewis/bridle:2002}. In the nominal cases we estimate the six \LCDM\ parameters: baryon density, $\Omega_b h^2$, cold dark matter density, $\Omega_c h^2$, acoustic peak angle, $\theta_A$ (reported in terms of $\theta_{\rm MC}$, an approximation of the acoustic peak angle that is used in CosmoMC) , amplitude, $A_s$ and scale dependence, $n_s$, of the primordial spectrum, defined at pivot scale $k=0.05$/Mpc, and optical depth to reionization, $\tau$. All have flat priors apart from the optical depth. We assume \neff$=3.046$ effective neutrino species, a Helium fraction of $Y_P=0.24$, a cosmological constant with $w=-1$, and following \planck\ \citep{planck_cosmo:2015} we fix the neutrino mass sum to $0.06$~eV.

\begin{figure}[t]
\includegraphics[width=\columnwidth]{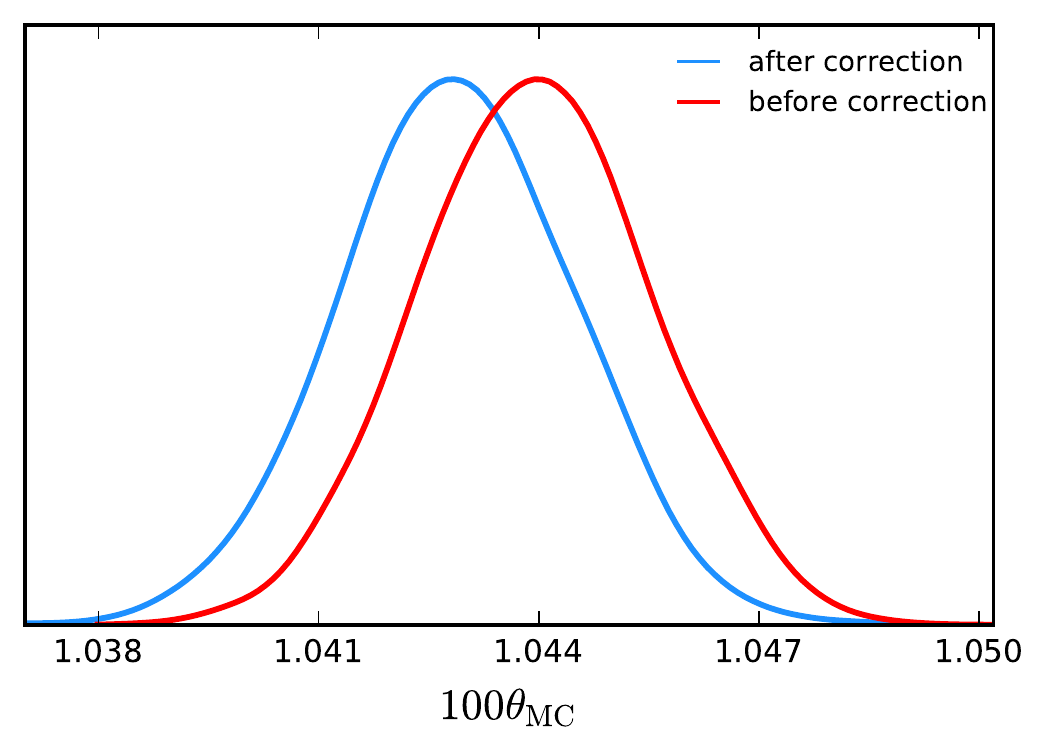}
\caption{Effect of aberration, due to our proper motion with respect to the CMB, on the peak position parameter $\theta$. The corrected power spectrum results in a $0.5\sigma$ decrease in the peak position.}
\label{fig:theta}
\end{figure}

We use the aberration-corrected spectra in our analysis, and test the effect on parameters with and without the correction. The ACTPol D56 patch is almost opposite to our direction of motion with respect to the last scattering surface.  An observer looking away from his or her direction of motion will measure the sound horizon to have a larger angular size compared to that seen by a comoving observer. As expected, we find a $0.5\sigma$ decrease in peak position $\theta$ when the correction is applied, as shown in Figure~\ref{fig:theta}. This effect must be accounted for when analyzing small regions of the sky; only over much larger regions does it average out for the two-point function.

\subsection{Goodness of fit of \LCDM}
\label{subsec:lcdm}

\begin{figure}[t!]
\includegraphics[width=\columnwidth]{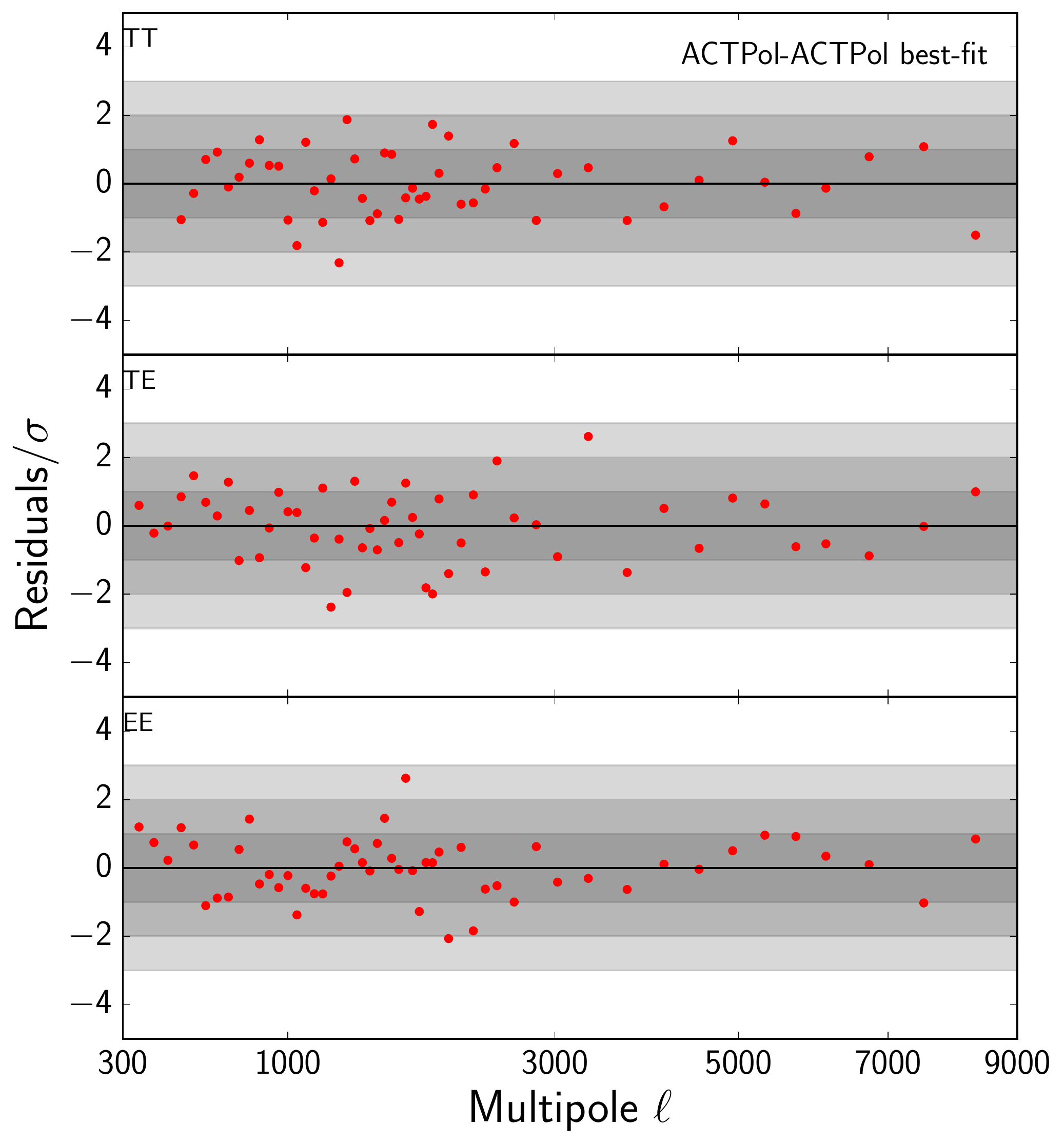}
\caption{The residuals between the ACTPol TT,TE, and EE power spectra and the best-fitting \LCDM\ model, in units of $\sigma$. The shaded bands show the 1,2 and 3$\sigma$ levels.}
\label{fig:lcdm_res}
\end{figure}

\begin{figure}[t!]
\includegraphics[width=\columnwidth]{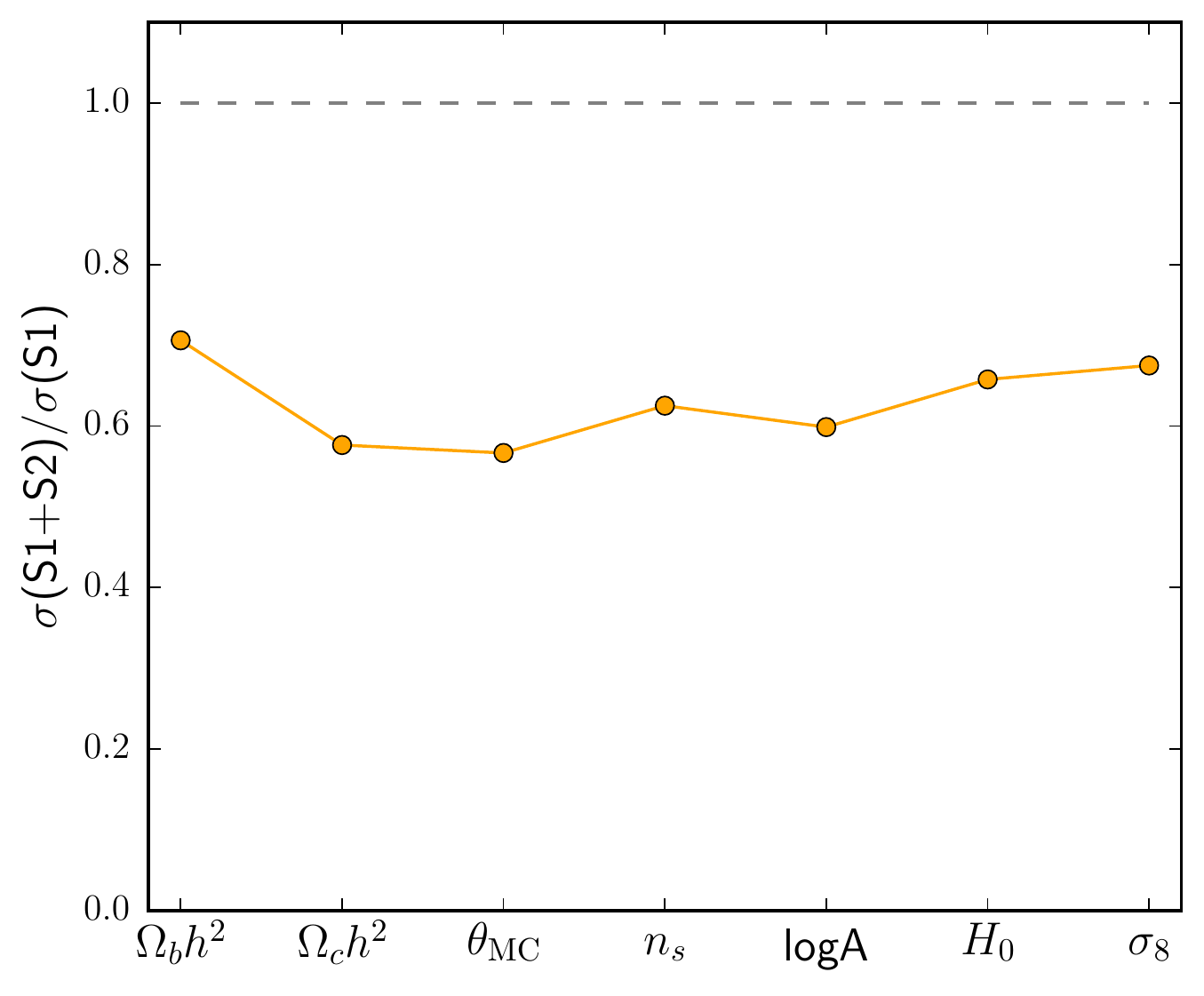}
\caption{The uncertainties in parameters estimated from ACTPol data are reduced from Season-1 to this Season-2 analysis to a factor of 0.6-0.7, gaining from increased observation time and wider sky coverage.}
\label{fig:s1s2}
\end{figure}

We first examine the best-fitting \LCDM\ model estimated using only ACTPol data. The  model is compared to the data in Figure~\ref{fig:lcdm_res}, where we show the residuals in standard deviations as a function of angular scale for TT, TE and EE.  This covers both the larger scales where the CMB dominates, and smaller scales where extragalactic foregrounds dominate in intensity. We do not find significant features beyond those expected due to noise. The reduced $\chi^2$ for this fit is 1.04 (for 142 degree of freedom). We find that the \LCDM\ model is an acceptable fit to the data.

\begin{figure*}[t]
  \includegraphics[width=\textwidth]{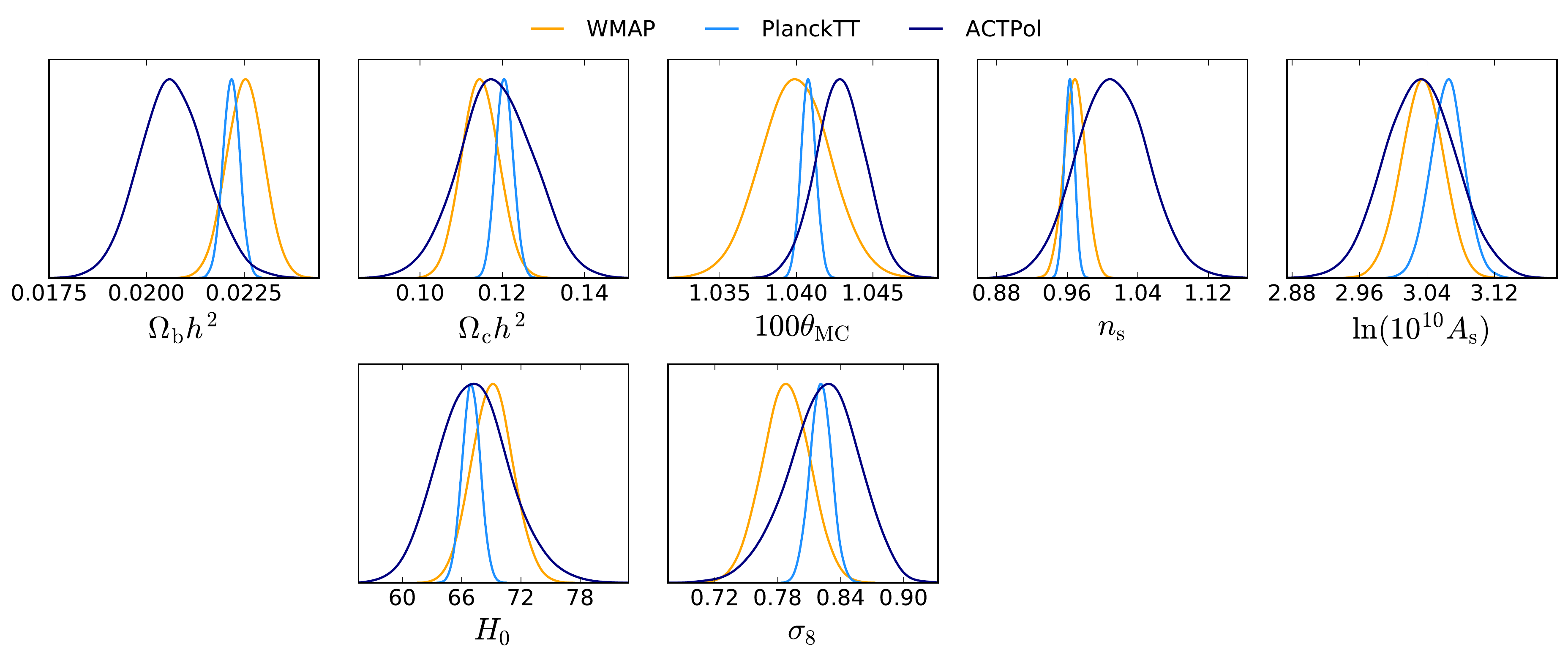}
\caption{Comparison of \LCDM\ parameters estimated from \wmap, \planck\ and ACTPol data. These likelihoods use 85\%, 66\%, and 1.4\% of the sky respectively.}
\label{fig:APall_lcdm}
\end{figure*}

The parameters estimated from the TT, TE and EE two-point functions  are shown in Figure~\ref{fig:APall_lcdm}. These are consistent with estimates from both \wmap\ and \planck, but would need to be combined with large-scale data to give competitive constraints. Despite not measuring the first acoustic peak, ACTPol data are able to constrain the peak position with higher precision than \wmap\ due to its measurement over a wide range of angular scales.

\subsection{Comparison to first-season data}
\label{subsec:lcdm}

Our second-season D56 data covers approximately twice the sky area observed in D1, D5 and D6 in S1. This reduction in cosmic variance uncertainty, together with the increase in observing time, translates into an improvement in cosmological parameters. In Figure~\ref{fig:s1s2}, we show the improvement between the Season-1 parameters derived from the \cite{naess/etal:2014} data, analyzed using the same priors as this analysis,  compared to the new data used in this paper. Estimates of the means are within 1-$\sigma$ for all parameters. The individual errors are reduced by a factor of between 1.4 and 1.7, corresponding to a ten-fold reduction in the five-dimensional parameter space volume.

\subsection{Relative contribution of temperature and polarization data}
\label{subsec:ttteee}

\begin{figure*}[thb!]
\includegraphics[width=\textwidth]{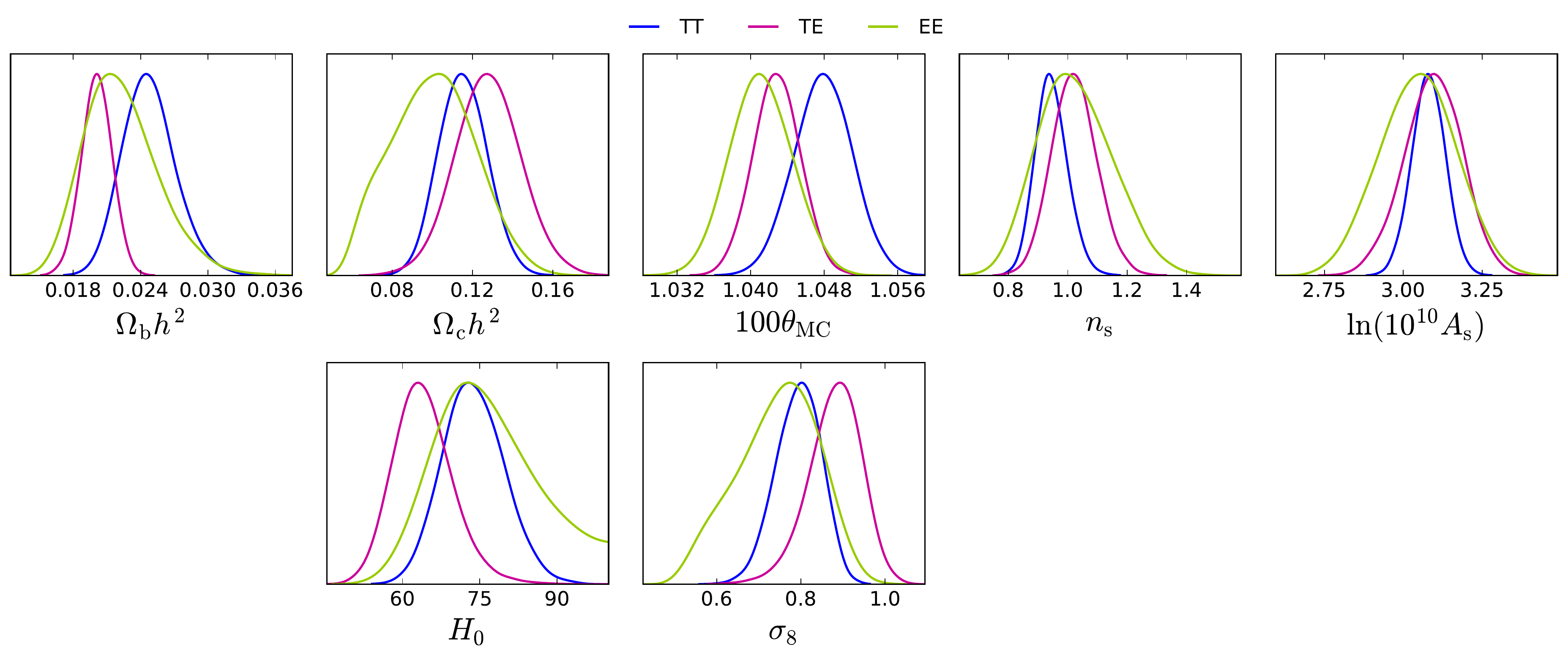}
\caption{\LCDM\ parameters as measured by different ACTPol spectra, sampled directly (top) and derived (bottom). The TE spectrum now provides the strongest internal ACTPol constraint on the baryon density, peak position, and Hubble constant.\\}
\label{fig:actpol_lcdm}
\end{figure*}

We then examine the relative contributions of the TT, TE and EE power spectrum in constraining the \LCDM\ model, and assess their consistency. We show parameters in Figure~\ref{fig:actpol_lcdm} and report constraints in Table~\ref{tab:acpol_lcdm}. We find good agreement for parameters derived from TT, TE and EE only spectra, and, for the first time, we find that multiple parameters are better constrained by the TE spectrum than the TT spectrum, using just the data measured by ACTPol.

The ACTPol TE spectrum now provides the tightest internal constraint on the baryon density and the peak position, compared to ACTPol TT and EE, and in turn provides the strongest internal constraint on the Hubble constant.  This strength of TE compared to TT was only marginally true for the data from \planck\ \citep{planck_cosmo:2015}, which had higher noise levels than ACTPol but mapped a larger region of the sky. There, the TE uncertainty on the CDM density was 0.95 the TT uncertainty, but all other \LCDM\ parameters were better constrained by TT.

Now, with ACTPol data, the error on the baryon density is 1.8 times smaller with TE than TT, and the peak position error is 1.3 times smaller. The EE spectrum is also starting to make an important contribution; for ACTPol the EE provides the same error on the peak position as the TT.

This is compatible with expectation, as discussed in e.g., \citet{galli/etal:2014}, that parameters which are constrained by the position and shape of the acoustic peaks get more weight from polarization data as the noise is further reduced. The peaks and troughs in the temperature power spectrum are less pronounced due to the contribution from the Doppler effect from velocity perturbations that are out of phase with the density perturbations. As a result, the peaks in the TT power spectrum have a lower contrast compared to the peaks in the polarization power spectrum, and the signal to noise on the location of the peaks and their amplitude is higher for polarization data.

In contrast, ACTPol parameters measured using the overall shape of the spectra are currently still better constrained by the temperature power spectrum, in particular the primordial amplitude $A_s$, because the signal to noise in the damping tail is higher for our two-season ACTPol temperature data.

\begin{table*}[ht!]
\centering
\caption{\small Comparison of \LCDM\ cosmological parameters and 68\% confidence intervals for ACTPol spectra. A Gaussian prior on the optical depth of $\tau=0.06\pm0.01$ is included.}
\begin{tabular}{ c  c  c  c c }
\hline
\hline
  & TT   &   TE & EE  & TT+TE+EE \\
& & & \\
 \hline
$100\Omega_bh^2$  & $2.47\pm0.23$   & $2.01\pm0.13$  & $2.23\pm0.34$ & $2.068\pm0.084$ \\
$100\Omega_ch^2$     & $11.5\pm1.2$  & $12.8\pm1.6$ & $10.0\pm2.0$ & $11.87\pm0.89$  \\
$10^4\theta_{MC}$     & $104.78\pm0.32$    & $104.27\pm0.25$  & $104.12\pm0.33$ & $104.29\pm0.16$  \\

$ln(10^{10}A_s)$ & $3.080\pm0.053$   & $3.096\pm0.090$  & $3.05\pm0.12$ & $3.032\pm0.041$  \\
$n_s$        & $0.947\pm0.053$   & $1.022\pm0.074$ & $1.03\pm0.12$ & $1.010\pm0.039$  \\
\hline
Derived &&&\\
$\sigma_8$   & $0.793\pm0.043 $   & $0.880\pm0.063$  & $0.742\pm0.094$ & $0.823\pm0.033$  \\ 
$H_0$        & $ 73.4\pm5.8 $ & $63.4\pm5.6$ & $76.7\pm9.4$  & $67.3\pm3.6$  \\
\hline
\end{tabular}
\label{tab:acpol_lcdm}
\end{table*}

\begin{figure*}[ht!]
\includegraphics[width=\textwidth]{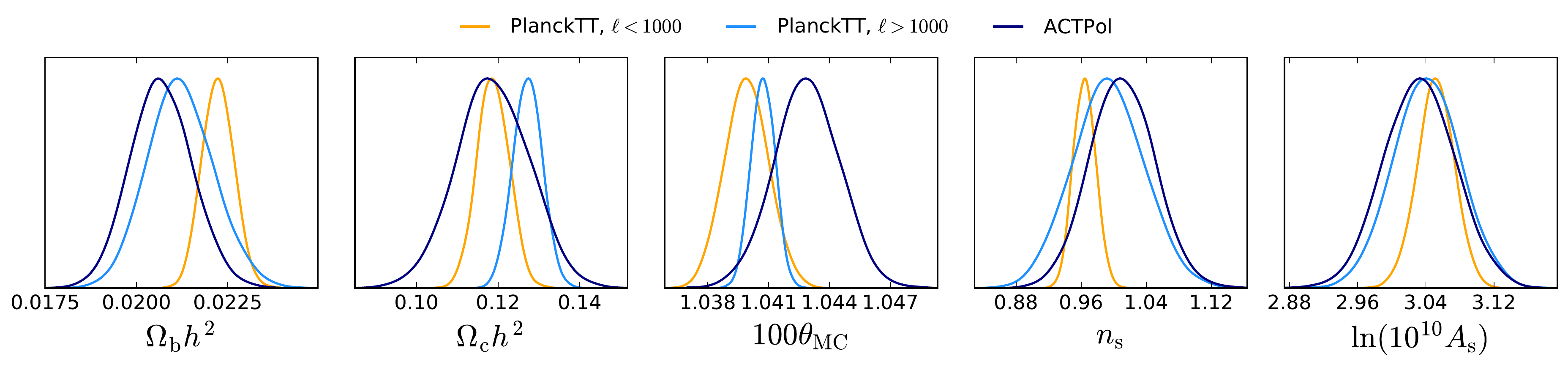}
\caption{Estimates of \LCDM\ parameters from ACTPol compared to parameters estimated from large and small multipole ranges of the \planck\ data. Current ACTPol data are consistent with both subsets of \planck. All models have a prior on the optical depth.\\}
\label{fig:ptest}
\end{figure*}

\begin{figure}[htb!]
\includegraphics[width=\columnwidth]{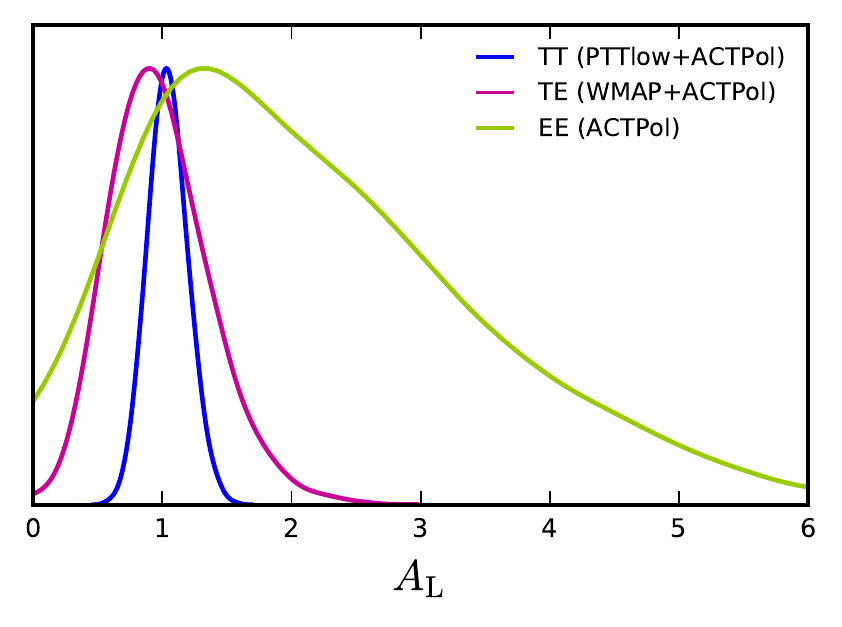}
\caption{Estimates of the lensing parameter $A_L$ using the TT, TE, and EE ACTPol data separately, combined with large-scale data.}
\label{fig:alens}
\end{figure}

\subsection{Consistency of TT and TE to \LCDM\ extensions}

Given the improved constraining power of TE, we explore whether any extensions of \LCDM\ are preferred by TE compared to TT. The TE spectrum offers an independent check of the model, and is not contaminated by emission from extragalactic foregrounds and SZ effects. As such, it is playing an increasingly important role in parameter constraints. 

We estimate the lensing parameter $A_L$, defined in \citealp{Calabrese2008}, through its effect on the smearing of the CMB acoustic peaks. To reduce degeneracy with other \LCDM\ parameters we add the \planck\ temperature and \wmap\ TE data at large scales, where the impact of lensing is minimal, and estimate $A_L$ jointly with the other \LCDM\ parameters. 

For the TT, TE, and EE data separately, we find marginalized distributions shown in Figure \ref{fig:alens}, with
\ba
A_L &=& 1.04 \pm 0.16 \quad {\rm TT \ (PTTlow+ACTPol)} \nonumber\\
A_L &=& 0.99 \pm 0.40 \quad {\rm TE \ (WMAP+ACTPol)} \nonumber\\
A_L &=& 2.1 \pm 1.3 \quad \ \ \ {\rm EE \ (ACTPol)} \,.
\ea
In all three cases we find that $A_L$ is consistent with the standard prediction of $A_L=1$. The TE power spectrum does not show signs of deviation from the expected lensing signal, and we now measure the lensing in the WMAP+ACT TE power spectrum at $2.5\sigma$ significance.

We repeat the same test with the number of relativistic species, and find  no evidence of deviation from the nominal $N_{\rm eff}=3.04$ in the TE or EE spectrum.

\subsection{Comparison to \planck}

Previous analyses of the \planck\ temperature data have shown a 2-3$\sigma$ difference in some parameters estimated from the small and large angular ranges of the \planck\ dataset \citep{2016ApJ...818..132A,2016arXiv160802487P}.
We compare parameters derived from our full ACTPol dataset to these two slicings of the \planck\ data.
In Figure~\ref{fig:ptest} we show parameters estimated from the ACTPol TT, TE and EE power spectra with parameters obtained from \planck\ temperature data using angular scales greater or smaller than $\ell=1000$. The ACTPol data presented in this paper are consistent with both sets of parameters estimated from \planck. Additional data from the third-season ACTPol observations will shed further light on this issue.

\subsection{Damping tail parameters}
\label{subsec:entend}

\begin{figure}
\includegraphics[width=8.5cm]{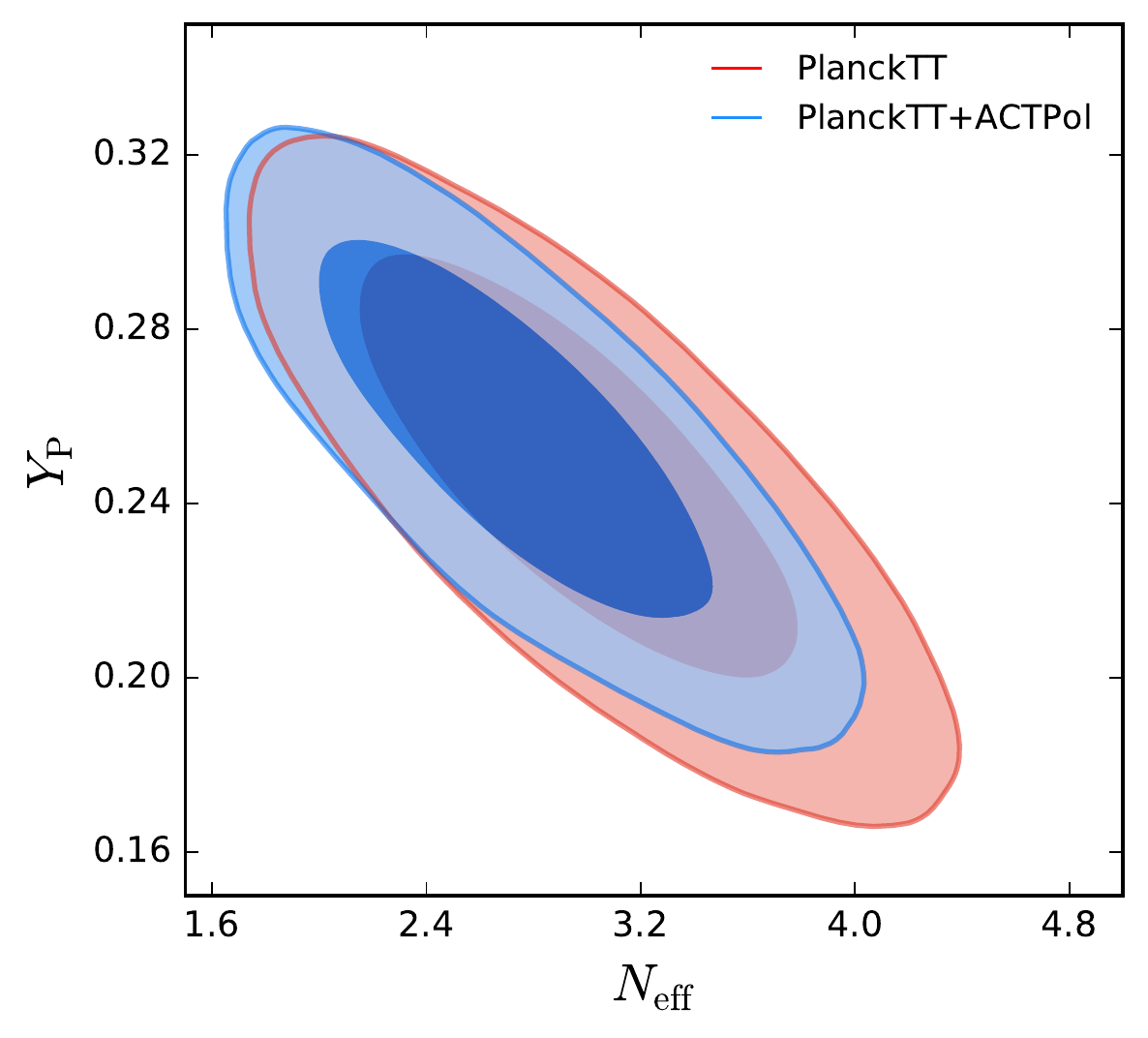}
\caption{Estimates of the number of relativistic species and primordial Helium abundance (68\% and 95\% CL) from \planck\ temperature data, and \planck\ combined with ACTPol.}
\label{fig:nnuyp}
\end{figure}

Given the consistency of the ACTPol data, both internally and with \planck, we add the ACTPol data to the full \planck\ temperature data to better constrain the effective number of relativistic species, and the primordial helium fraction.  Figure~\ref{fig:nnuyp} shows the improvement on the 68\% and 95\% confidence levels by adding the ACTPol data to the \planck\ temperature data ($2<\ell<2500$).
We find
\ba
N_{\rm eff} =  2.74 &\pm&  \rm 0.47  \nonumber\\
Y_{\rm P}=  0.255&\pm&  \rm 0.027 \quad (PlanckTT+ACTPol) 
\ea
compared to $N_{\rm eff}= 2.99 \pm 0.52$ and $Y_P = 0.246 \pm 0.031$ from PlanckTT alone.

Additional ACTPol data measuring the damping tail data will further tighten these limits and better test the standard paradigm.

%% file: conclude.tex
\section{Conclusions}
\label{sec:conc}

We have presented temperature and polarization power spectra estimated from 548 deg$^2$ of sky observed at night during the first two seasons of ACTPol observation. We find good agreement between cosmological parameters estimated from the TT, TE and EE power spectra individually, and the spectra are consistent with the \LCDM\ model. The CMB temperature-polarization correlation is now more constraining than the temperature anisotropy for certain parameters; the baryon density and acoustic peak angular scale are now best internally constrained from the TE power spectrum. Adding the new ACTPol polarization data to the \planck\ temperature data improves constraints on extensions to the \LCDM\ model that affect the damping tail.    

This analysis includes only 12\% of the full three-season ACTPol data taken from 2013--15, so future analyses will provide an opportunity to further test the \LCDM\ model and more tightly constrain properties including the number of relativistic species, the sum of neutrino masses, and the primordial power spectrum.

%% file: spectra_tables.tex
Table \ref{tab:spectra} shows the combined 149 GHz spectra from D5, D6 and D56, and Table \ref{tab:cmb_spectra} shows the foreground marginalized
CMB spectra used for cosmological parameter estimation. The spectra and likelihood presented in this paper are available on LAMBDA: https://lambda.gsfc.nasa.gov/product/act/
%
\begin{table*}[htb!]\small
\caption{ ACTPol Two-season night-time Power Spectra in D56 region, ${\cal{D}}_\ell = \ell(\ell+1)C_{\ell}/2\pi$ ($\mu$K$^2$).}
\vspace{-0.2in}
\begin{center}
\begin{tabular}{ c c  | r  r |  r  r | r  r | r  r | r  r | r  r }
\hline
\hline
$\ell$ & $\ell$ range & \multicolumn{2}{c|}{TT} & \multicolumn{2}{c|}{TE} & \multicolumn{2}{c|}{EE} & \multicolumn{2}{c|}{BB} & \multicolumn{2}{c|}{TB} & \multicolumn{2}{c}{EB} \\
  & & ${\cal{D}}_\ell$ & $\sigma ({\cal{D}}_\ell)$
& ${\cal{D}}_\ell$ & $\sigma ({\cal{D}}_\ell)$
& ${\cal{D}}_\ell$ & $\sigma ({\cal{D}}_\ell)$
& ${\cal{D}}_\ell$ & $\sigma ({\cal{D}}_\ell)$
& ${\cal{D}}_\ell$ & $\sigma ({\cal{D}}_\ell)$
& ${\cal{D}}_\ell$ & $\sigma ({\cal{D}}_\ell)$
\\
\hline 
 350 & $ 325- 375$ & 2452.3 &  299.5 &   85.7 &   19.2 &   18.8 &    1.8 &    0.3 &    0.3 &  -10.0 &    7.9 &    0.2 &    0.5 \\
 400 & $ 375- 425$ & 1582.8 &  209.1 &   -9.5 &   15.7 &   21.4 &    2.0 &   -0.1 &    0.3 &    0.3 &    6.2 &    0.8 &    0.5 \\
 450 & $ 425- 475$ & 2005.1 &  187.9 &  -65.8 &   14.1 &   15.5 &    1.5 &    0.1 &    0.3 &   -5.0 &    6.3 &   -0.6 &    0.4 \\
 500 & $ 475- 525$ & 2219.3 &  208.4 &  -44.0 &   12.6 &    9.5 &    1.0 &    0.0 &    0.3 &    1.7 &    7.3 &   -0.1 &    0.4 \\
 550 & $ 525- 575$ & 2477.7 &  194.9 &   16.7 &   11.9 &    9.5 &    1.1 &    0.2 &    0.3 &   -4.2 &    6.9 &    0.0 &    0.4 \\
 600 & $ 575- 625$ & 2298.5 &  153.1 &   32.6 &   13.7 &   17.7 &    1.8 &    0.3 &    0.3 &    0.5 &    6.3 &   -0.1 &    0.5 \\
 650 & $ 625- 675$ & 1985.6 &  134.7 &  -16.9 &   14.9 &   29.9 &    2.6 &   -0.4 &    0.4 &    6.2 &    5.9 &    0.1 &    0.7 \\
 700 & $ 675- 725$ & 1878.8 &  126.0 &  -76.7 &   14.7 &   33.0 &    2.8 &    0.1 &    0.4 &   -2.4 &    5.7 &    0.3 &    0.7 \\
 750 & $ 725- 775$ & 2251.6 &  135.7 & -143.0 &   14.4 &   27.9 &    2.2 &   -0.1 &    0.4 &   -2.6 &    6.3 &    0.7 &    0.6 \\
 800 & $ 775- 825$ & 2572.6 &  144.0 &  -80.0 &   11.4 &   17.7 &    1.5 &    0.7 &    0.4 &    3.8 &    6.5 &   -0.0 &    0.5 \\
 850 & $ 825- 875$ & 2529.5 &  138.1 &  -13.1 &   10.8 &   13.0 &    1.4 &    0.3 &    0.4 &   -1.3 &    6.3 &    0.9 &    0.5 \\
 900 & $ 875- 925$ & 1910.9 &   96.0 &   47.3 &   10.5 &   23.3 &    1.9 &   -0.1 &    0.4 &   -0.3 &    5.4 &   -0.9 &    0.6 \\
 950 & $ 925- 975$ & 1381.6 &   74.2 &   40.7 &   10.8 &   34.9 &    2.7 &    0.5 &    0.5 &   -9.1 &    4.5 &    0.3 &    0.8 \\
1000 & $ 975-1025$ & 1025.9 &   62.2 &  -24.8 &    9.9 &   39.5 &    2.7 &   -0.1 &    0.5 &   11.6 &    4.1 &    0.8 &    0.8 \\
1050 & $1025-1075$ & 1012.3 &   59.7 &  -69.6 &    9.1 &   28.7 &    2.3 &    0.2 &    0.5 &    0.4 &    4.3 &    0.4 &    0.7 \\
1100 & $1075-1125$ & 1307.1 &   61.0 &  -78.8 &    8.5 &   18.4 &    1.7 &    0.2 &    0.6 &   -6.7 &    4.7 &   -0.0 &    0.7 \\
1150 & $1125-1175$ & 1211.2 &   59.3 &  -33.0 &    7.2 &   12.5 &    1.4 &   -0.1 &    0.6 &    0.3 &    4.5 &    1.1 &    0.7 \\
1200 & $1175-1225$ & 1000.0 &   51.1 &    7.4 &    7.2 &   16.1 &    1.6 &    1.0 &    0.6 &    2.8 &    4.2 &   -0.7 &    0.7 \\
1250 & $1225-1275$ &  858.6 &   38.3 &  -21.6 &    7.1 &   24.9 &    2.1 &    0.9 &    0.6 &    5.2 &    3.7 &   -0.1 &    0.8 \\
1300 & $1275-1325$ &  663.8 &   36.3 &  -37.3 &    7.1 &   29.4 &    2.2 &    0.1 &    0.7 &    0.3 &    3.7 &    0.5 &    0.8 \\
1350 & $1325-1375$ &  835.4 &   36.4 &  -71.2 &    6.8 &   26.8 &    2.2 &    0.8 &    0.8 &    2.0 &    3.8 &   -0.5 &    0.9 \\
1400 & $1375-1425$ &  846.7 &   38.1 &  -45.6 &    6.3 &   18.0 &    1.7 &    0.7 &    0.7 &   -4.8 &    3.9 &    0.2 &    0.7 \\
1450 & $1425-1475$ &  785.0 &   36.0 &  -31.4 &    5.7 &   12.0 &    1.4 &   -0.3 &    0.8 &   -5.8 &    3.9 &    0.3 &    0.7 \\
1500 & $1475-1525$ &  656.0 &   29.6 &   -3.6 &    5.1 &   12.5 &    1.4 &    1.7 &    0.8 &   -1.6 &    3.4 &    0.3 &    0.7 \\
1550 & $1525-1575$ &  521.0 &   24.1 &   -1.5 &    5.1 &   18.2 &    1.7 &    0.9 &    0.9 &    0.0 &    3.3 &    0.4 &    0.8 \\
1600 & $1575-1625$ &  456.4 &   20.0 &   -9.4 &    4.9 &   22.6 &    2.0 &   -0.5 &    0.9 &    3.0 &    3.1 &   -0.9 &    0.9 \\
1650 & $1625-1675$ &  420.8 &   19.3 &  -22.0 &    4.6 &   18.6 &    1.8 &    0.1 &    0.9 &    0.3 &    3.0 &   -1.5 &    0.9 \\
1700 & $1675-1725$ &  389.4 &   18.5 &  -31.6 &    4.2 &   13.3 &    1.6 &    0.7 &    0.9 &    0.3 &    2.9 &   -0.8 &    0.8 \\
1750 & $1725-1775$ &  396.2 &   17.4 &  -16.6 &    3.9 &   12.9 &    1.4 &   -0.2 &    0.9 &    1.8 &    3.1 &   -0.3 &    0.8 \\
1800 & $1775-1825$ &  363.5 &   16.1 &   -9.8 &    3.8 &    7.8 &    1.4 &    0.5 &    1.0 &    6.1 &    2.9 &   -0.1 &    0.8 \\
1850 & $1825-1875$ &  303.2 &   14.3 &   -7.2 &    3.7 &    7.3 &    1.6 &   -1.0 &    1.0 &   -0.2 &    2.7 &   -0.2 &    0.8 \\
1900 & $1875-1925$ &  261.0 &   12.5 &  -16.7 &    3.7 &   11.0 &    1.6 &   -2.0 &    1.1 &    0.0 &    2.7 &    0.2 &    1.0 \\
1950 & $1925-1975$ &  267.2 &   11.8 &  -22.9 &    3.5 &   10.7 &    1.7 &   -1.2 &    1.1 &   -1.0 &    2.7 &   -1.7 &    1.0 \\
2000 & $1975-2025$ &  247.9 &   11.8 &  -15.3 &    3.5 &    9.2 &    1.7 &    1.9 &    1.2 &    2.3 &    2.6 &    0.9 &    1.0 \\
2075 & $2025-2125$ &  237.5 &    7.5 &  -14.3 &    2.3 &    3.6 &    1.0 &    0.3 &    0.9 &    0.9 &    1.8 &    0.3 &    0.6 \\
2175 & $2125-2225$ &  166.1 &    6.1 &   -5.1 &    2.0 &    6.4 &    1.2 &   -1.0 &    0.9 &   -0.7 &    1.6 &   -0.6 &    0.7 \\
2275 & $2225-2325$ &  130.4 &    5.0 &   -5.5 &    2.0 &    3.5 &    1.2 &    0.6 &    1.0 &   -0.8 &    1.5 &    0.3 &    0.7 \\
2375 & $2325-2425$ &  120.0 &    4.8 &   -9.1 &    1.8 &    3.0 &    1.2 &    0.5 &    1.1 &   -3.9 &    1.6 &    1.1 &    0.7 \\
2475 & $2425-2525$ &  101.3 &    4.2 &   -0.3 &    1.7 &    2.5 &    1.2 &    0.7 &    1.1 &    2.3 &    1.5 &   -1.4 &    0.8 \\
2625 & $2525-2725$ &   81.4 &    2.5 &   -3.6 &    1.2 &    1.8 &    0.9 &    0.1 &    0.9 &    0.4 &    1.0 &    1.0 &    0.6 \\
2825 & $2725-2925$ &   55.4 &    2.1 &   -1.9 &    1.1 &    2.6 &    1.0 &   -1.3 &    1.0 &   -0.7 &    0.9 &   -0.8 &    0.7 \\
3025 & $2925-3125$ &   46.2 &    1.8 &   -2.5 &    1.1 &    1.1 &    1.1 &   -0.3 &    1.1 &   -0.6 &    0.9 &    0.3 &    0.8 \\
3325 & $3125-3525$ &   36.9 &    1.2 &    1.3 &    0.8 &    1.1 &    0.9 &    1.5 &    0.9 &   -0.6 &    0.7 &   -0.7 &    0.6 \\
3725 & $3525-3925$ &   30.7 &    1.2 &   -1.5 &    0.9 &    0.6 &    1.2 &    0.1 &    1.2 &   -0.1 &    0.8 &   -0.1 &    0.8 \\
4125 & $3925-4325$ &   31.8 &    1.3 &    0.3 &    1.1 &    1.6 &    1.5 &    1.8 &    1.4 &    0.4 &    0.9 &    1.1 &    1.0 \\
4525 & $4325-4725$ &   35.7 &    1.5 &   -1.0 &    1.3 &    1.6 &    1.9 &    2.1 &    1.9 &    2.5 &    1.1 &   -0.1 &    1.2 \\
4925 & $4725-5125$ &   41.9 &    1.8 &    1.1 &    1.6 &    3.1 &    2.3 &   -1.6 &    2.2 &   -3.4 &    1.3 &   -1.6 &    1.6 \\
5325 & $5125-5525$ &   44.5 &    2.1 &    1.0 &    1.8 &    4.9 &    2.8 &    4.2 &    2.9 &   -0.7 &    1.6 &   -0.7 &    1.9 \\
5725 & $5525-5925$ &   47.7 &    2.4 &   -1.6 &    2.3 &    5.8 &    3.4 &   -1.0 &    3.5 &    2.9 &    2.0 &    3.5 &    2.4 \\
6125 & $5925-6325$ &   55.0 &    2.9 &   -1.7 &    2.8 &    4.5 &    4.4 &   -0.3 &    4.5 &   -1.0 &    2.4 &    0.8 &    3.0 \\
6725 & $6325-7125$ &   67.2 &    2.8 &   -2.5 &    2.6 &    4.0 &    4.0 &    6.9 &    4.1 &    1.4 &    2.2 &   -1.9 &    2.8 \\
7525 & $7125-7925$ &   83.8 &    4.1 &   -0.4 &    3.9 &   -2.2 &    6.5 &    3.8 &    6.4 &    0.4 &    3.5 &   -9.5 &    4.3 \\
8325 & $7925-8725$ &   86.9 &    6.4 &    5.8 &    6.2 &   14.1 &   10.1 &   19.2 &    9.6 &    0.1 &    5.1 &    1.2 &    7.0 \\
\hline
\end{tabular}
\end{center}
\label{tab:spectra}
\end{table*}

\begin{table*}[htb!]\small
\caption{ ACTPol Two-season foregound-marginalized Power Spectra in D56 region, ${\cal{D}}_\ell = \ell(\ell+1)C_{\ell}/2\pi$ ($\mu$K$^2$).}
\vspace{-0.2in}
\begin{center}
\begin{tabular}{ c c  | r  r |  r  r | r  r }
\hline
\hline
$\ell$ & $\ell$ range & \multicolumn{2}{c|}{TT} & \multicolumn{2}{c|}{TE} & \multicolumn{2}{c}{EE} \\
 & & ${\cal{D}}_\ell$ & $\sigma ({\cal{D}}_\ell)$
& ${\cal{D}}_\ell$ & $\sigma ({\cal{D}}_\ell)$
& ${\cal{D}}_\ell$ & $\sigma ({\cal{D}}_\ell)$
\\
\hline
 350 & $ 325- 375$ &    -- &    -- &   85.5 &   19.1 &   18.8 &    1.8 \\
 400 & $ 375- 425$ &    -- &    -- &   -9.4 &   15.7 &   21.3 &    2.0 \\
 450 & $ 425- 475$ &  -- & -- &  -65.7 &   14.2 &   15.4 &    1.5 \\
 500 & $ 475- 525$ & 2211.2 &  208.4 &  -43.9 &   12.5 &    9.5 &    1.0 \\
 550 & $ 525- 575$ & 2468.0 &  194.7 &   16.7 &   11.9 &    9.4 &    1.1 \\
 600 & $ 575- 625$ & 2289.3 &  153.1 &   32.6 &   13.6 &   17.6 &    1.8 \\
 650 & $ 625- 675$ & 1977.6 &  134.3 &  -16.8 &   14.9 &   29.8 &    2.6 \\
 700 & $ 675- 725$ & 1870.7 &  125.7 &  -76.6 &   14.7 &   32.9 &    2.8 \\
 750 & $ 725- 775$ & 2242.3 &  135.5 & -142.7 &   14.4 &   27.8 &    2.2 \\
 800 & $ 775- 825$ & 2564.4 &  143.7 &  -80.0 &   11.4 &   17.6 &    1.5 \\
 850 & $ 825- 875$ & 2521.5 &  137.8 &  -13.0 &   10.8 &   12.9 &    1.4 \\
 900 & $ 875- 925$ & 1902.2 &   95.7 &   47.3 &   10.5 &   23.2 &    1.9 \\
 950 & $ 925- 975$ & 1373.1 &   73.8 &   40.7 &   10.8 &   34.7 &    2.6 \\
1000 & $ 975-1025$ & 1017.2 &   62.0 &  -24.7 &    9.8 &   39.3 &    2.7 \\
1050 & $1025-1075$ & 1003.7 &   59.6 &  -69.6 &    9.1 &   28.5 &    2.3 \\
1100 & $1075-1125$ & 1298.3 &   60.8 &  -78.8 &    8.5 &   18.3 &    1.7 \\
1150 & $1125-1175$ & 1202.2 &   59.0 &  -33.0 &    7.2 &   12.3 &    1.4 \\
1200 & $1175-1225$ &  990.8 &   51.1 &    7.4 &    7.2 &   15.9 &    1.6 \\
1250 & $1225-1275$ &  849.4 &   38.1 &  -21.5 &    7.1 &   24.8 &    2.0 \\
1300 & $1275-1325$ &  654.2 &   36.3 &  -37.3 &    7.1 &   29.1 &    2.2 \\
1350 & $1325-1375$ &  825.5 &   36.4 &  -71.1 &    6.8 &   26.6 &    2.2 \\
1400 & $1375-1425$ &  836.3 &   38.1 &  -45.6 &    6.2 &   17.8 &    1.7 \\
1450 & $1425-1475$ &  774.6 &   36.1 &  -31.4 &    5.7 &   11.7 &    1.4 \\
1500 & $1475-1525$ &  645.4 &   29.6 &   -3.6 &    5.1 &   12.3 &    1.4 \\
1550 & $1525-1575$ &  510.0 &   24.1 &   -1.4 &    5.1 &   17.9 &    1.7 \\
1600 & $1575-1625$ &  445.0 &   20.0 &   -9.4 &    4.9 &   22.3 &    1.9 \\
1650 & $1625-1675$ &  408.9 &   19.4 &  -22.0 &    4.6 &   18.3 &    1.8 \\
1700 & $1675-1725$ &  377.5 &   18.5 &  -31.5 &    4.2 &   12.9 &    1.6 \\
1750 & $1725-1775$ &  383.9 &   17.3 &  -16.6 &    3.9 &   12.5 &    1.4 \\
1800 & $1775-1825$ &  351.0 &   16.1 &   -9.8 &    3.8 &    7.5 &    1.4 \\
1850 & $1825-1875$ &  290.4 &   14.3 &   -7.2 &    3.7 &    6.9 &    1.6 \\
1900 & $1875-1925$ &  247.9 &   12.5 &  -16.7 &    3.8 &   10.5 &    1.6 \\
1950 & $1925-1975$ &  253.8 &   11.8 &  -22.8 &    3.5 &   10.3 &    1.7 \\
2000 & $1975-2025$ &  234.2 &   11.8 &  -15.2 &    3.5 &    8.8 &    1.7 \\
2075 & $2025-2125$ &  223.3 &    7.5 &  -14.3 &    2.2 &    3.1 &    1.0 \\
2175 & $2125-2225$ &  151.2 &    6.1 &   -5.1 &    2.0 &    5.8 &    1.2 \\
2275 & $2225-2325$ &  114.8 &    5.1 &   -5.4 &    2.0 &    2.9 &    1.2 \\
2375 & $2325-2425$ &  103.7 &    4.9 &   -9.0 &    1.8 &    2.3 &    1.2 \\
2475 & $2425-2525$ &   84.2 &    4.2 &   -0.2 &    1.7 &    1.8 &    1.2 \\
2625 & $2525-2725$ &   63.3 &    2.6 &   -3.6 &    1.2 &    1.1 &    0.9 \\
2825 & $2725-2925$ &   35.7 &    2.3 &   -1.8 &    1.1 &    1.7 &    1.1 \\
3025 & $2925-3125$ &   24.9 &    2.0 &   -2.4 &    1.1 &    0.1 &    1.2 \\
3325 & $3125-3525$ &   13.1 &    1.5 &    1.4 &    0.8 &   -0.2 &    1.0 \\
3725 & $3525-3925$ &    3.2 &    1.4 &   -1.4 &    0.9 &   -1.0 &    1.3 \\
4125 & $3925-4325$ &    0.4 &    1.5 &    0.5 &    1.2 &   -0.3 &    1.6 \\
\hline
\end{tabular}
\end{center}
\label{tab:cmb_spectra}
\end{table*}